\documentclass[12pt]{article}

\usepackage{times}
\usepackage{epsfig}
\usepackage{natbib}
\usepackage{graphicx,color}
\usepackage[fleqn]{amsmath}
\usepackage{amssymb}
\usepackage{multirow}
\usepackage{bbm}
\usepackage{mathtools}
\usepackage{authblk}
\usepackage{booktabs}
\usepackage{relsize}
\usepackage{setspace}

\usepackage[letterpaper, margin=1in]{geometry}
          
\newif\ifblind
\newif\ifunblind

\unblindtrue

\title{Nonparametric Bayes Models of Fiber Curves Connecting Brain Regions}
\ifblind
\author{author here}
\else
\author[1]{\smaller{Zhengwu Zhang} }
\author[1]{{Maxime Descoteaux } }
\author[2]{{David B. Dunson}}
\affil[1]{Department of Statistical Science, Duke University, United States}
\affil[2]{Computer Science Department, Faculty of Science, University of Sherbrooke, Canada}
\fi
\date{}

\def\argmin{\mathop{\rm argmin}}

\newcommand{\real}{\ensuremath{\mathbb{R}}}

\def\argmin{\mathop{\rm argmin}}

\def \Y {\mathcal{Y}}
\def \K {\mathcal{K}}

\def \bO {{\bf{O}}}

\newcommand{\ra}[1]{\renewcommand{\arraystretch}{#1}}

\begin{document}

\maketitle
\onehalfspacing

\begin{abstract}
In studying structural inter-connections in the human brain, it is common to first estimate fiber bundles connecting different regions of the brain relying on diffusion MRI.  These fiber bundles act as highways for neural activity and communication, snaking through the brain and connecting different regions.  Current statistical methods for analyzing these fibers reduce the rich information into an adjacency matrix, with the elements containing a count of the number of fibers or a mean diffusion feature (such as fractional anisotropy) along the fibers.  The goal of this article is to avoid discarding the rich functional data on the shape, size and orientation of fibers, developing flexible models for characterizing the population distribution of fibers between brain regions of interest within and across different individuals.  We start by decomposing each fiber in each individual's brain into a corresponding rotation matrix, shape and translation from a global reference curve.  These components can be viewed as data lying on a product space composed of different Euclidean spaces and manifolds.  To non-parametrically model the distribution within and across individuals, we rely on a hierarchical mixture of product kernels specific to the component spaces.  Taking a Bayesian approach to inference, we develop an efficient method for posterior sampling.  The approach automatically produces clusters of fibers within and across individuals, and yields interesting new insights into variation in fiber curves, while providing a useful starting point for more elaborate models relating fibers to covariates and neuropsychiatric traits.

\end{abstract}

\newpage

\section{Introduction}

\begin{figure}
\begin{center}
\includegraphics[height=2.8in]{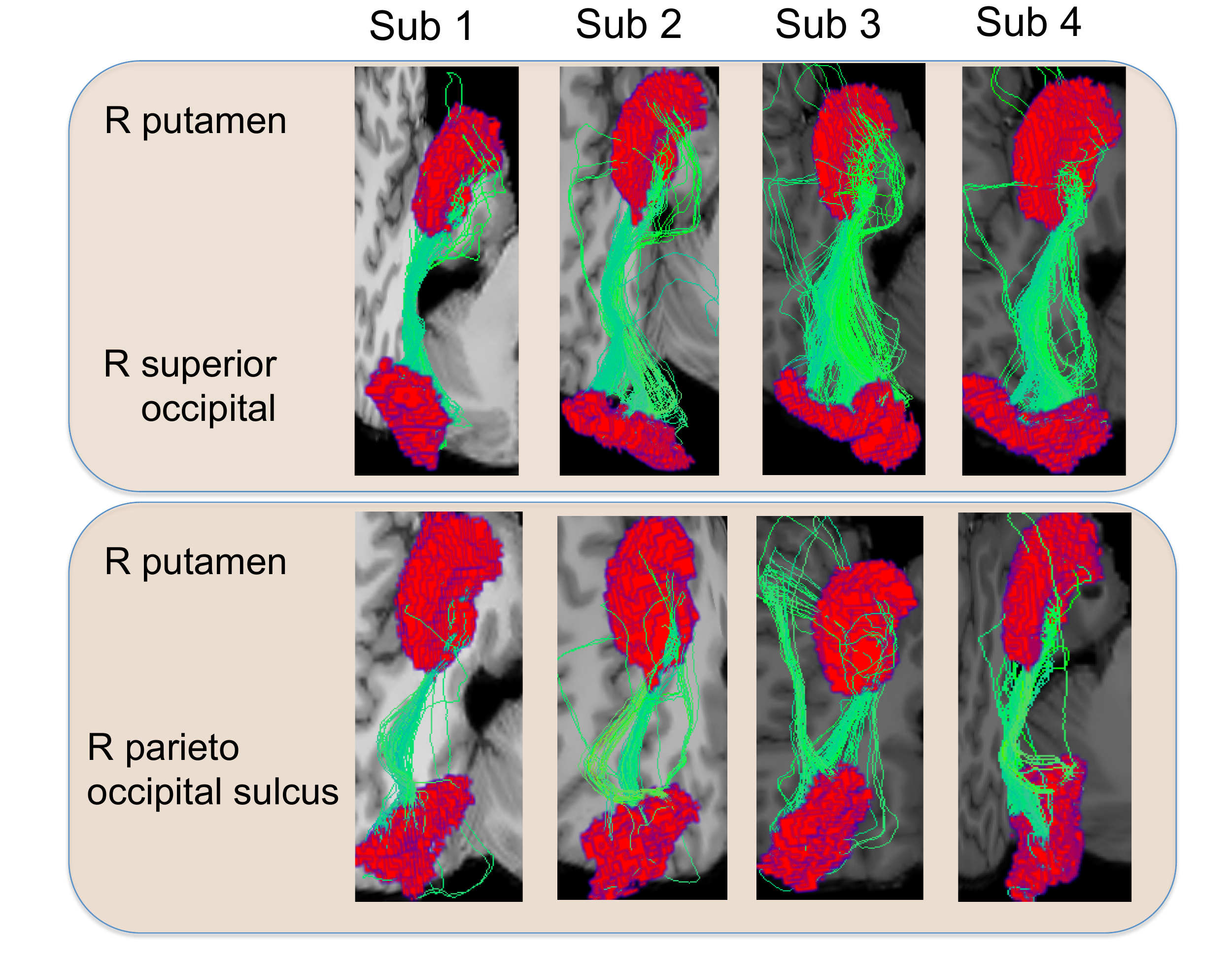} 
\caption{Examples of fiber curves connecting two different pairs of regions in four subjects. } \label{fig:fiber_ill1}
\end{center}
\end{figure}

There has been dramatically increasing interest in recent years in {\em connectomics}, which is the study of functional and structural interconnection networks in the human brain \citep{Jbabdi2015,Glasser2016,Park2013,Fornito2013}.  This interest has been spurred by the development of new imaging technologies, which allow researchers to non-invasively peer into the human brain and obtain data on connections.  The focus of this article is on {\em structural connections}, corresponding to fiber bundles that are estimated from diffusion magnetic resonance imaging (dMRI).   dMRI measures the diffusion of water molecules across tissues in the brain; this diffusion tends to be directional along white matter tracts acting as highways for neural activity, while being weaker and non-directional in gray matter.  By combining data from diffusion MRI and structural MRI, the brain can be segmented into different functional regions, with the fiber bundles connecting the different regions estimated.  Focusing on two regions of interest (ROIs) and applying a recent fiber tracking algorithm \citep{Smith2012,Girard2014266}, Figure \ref{fig:fiber_ill1} shows the fiber connections for four different individuals.  There are large numbers of fibers connecting these two ROIs, and there are interesting similarities and differences among the subjects in the fiber locations and shapes.

Fiber connections in each individual's brain can be viewed as a type of {\em object} data.  There are  many exciting possibilities in terms of relating these objects to traits of the individual.  For example, Figure 1 shows evidence of functional data clusters in the fiber connections, with the locations and numbers of fibers occupying these clusters varying across individuals; perhaps features of the clusters relate to traits of the individual, such as their intelligence or whether they have frequent migraines or episodes of depression.  Our over-arching goal is to develop the statistical and computational tools necessary to make such inferences. However, the current literature on statistical analysis of fiber tracts reduces the rich functional data to simple summary statistics prior to analysis.  In particular, current connectome pipelines output an adjacency matrix consisting of a count of the number of fibers in each pair of brain regions \citep{Reus2013,Fornito2013}, reducing the rich data shown in Figure 1 to a single count for each panel, discarding information on fiber shapes, sizes and locations.  These adjacency matrices are typically reduced further to a binary form \citep{Durante2014n,Durante2014} or to topological features of the network \citep{Cheng2012,Fornito2013} in order to simplify analyses of brain structure and its relationship with other factors.

Clearly, the data represented in Figure 1 are {\em functional data}, and hence it is natural to think of applying functional data analysis (FDA) methodology.  However, most FDA methods are developed for much simpler cases in which there is a single function $y_i: \mathcal{T} \to \Re$ for each individual, with $\mathcal{T} \subset \Re$.  For example, $y_i$ may represent a growth curve with age for individual $i$.  There is also a rich literature on more elaborate FDA models for curve data, allowing multivariate, hierarchical, spatial and temporal dependence structures \citep{Wang2015}.  Even in more complex cases, the majority of the focus has been on one-dimensional curves $y_i: \mathcal{T} \to \Re$, using a rich variety of representations ranging from spline expansions to functional principal components analysis (FPCA) to Gaussian process-based models.  After defining a representation for $y_i$, it is often relatively straightforward to further include structured dependence in the curves (e.g., hierarchical, spatial, temporal, etc).

The fiber tracts illustrated in Figure 1 are quite complex in corresponding to many three-dimensional curves snaking through $\Re^3$ having different intersection points with two non-regularly shaped brain regions of interest.  In addition, there is clear clustering evidence and heterogeneity among individuals.  It is not obvious how to define a model for these data, which is sufficiently flexible and captures the important characteristics, such as the clusters, without discarding too much information or becoming computationally intractable given the number of fibers.  There is a rich literature on nonparametric Bayesian models for functional data, which induce clustering \citep{Abel2009} and can even allow joint modeling of functional predictors with a response \citep{Jamie2009}, but these methods focus on the case in which a single function $y_i$ is observed for each individual, and hence are not directly relevant.

In this article, we propose a novel approach, which relies on characterizing each fiber curve within each individual in terms of its rotation, shape and translation from a global reference curve.  This allows us to define a nonparametric model for the fiber curve data through a dual representation of the data on a product space.  On this product space we define a mixture of product kernels motivated by the framework of \citet{Bhattacharya2012,Bhattacharya2010}, who showed conditions under which Dirichlet process mixtures of product kernels having support on different manifolds lead to consistent estimation of an unknown joint distribution of data on a product manifold.  However, their framework is abstract and they did not consider data consisting of rotation matrices or allow nested dependence, as we obtain due to nesting of the fibers within each individual's brain.

Section 2 describes the basic data structure and representation of fiber curves.  Section 3 proposes a product mixture model for fiber connections in an individual's brain, and outlines a Markov chain Monte Carlo (MCMC) algorithm for posterior inference. Section 4 proposes a nested Dirichlet process model for modeling fiber curves for a population of individuals.  Section 5 summarizes analyses of human brain connectomics data, and Section 6 discusses the results and outlines interesting next directions. 

\section{Fiber curves extraction and representation}
\subsection{Data description}
We use a state-of-the-art tractography algorithm \citep{Smith2012,Girard2014266} to generate the fiber tracts relying on two steps. 
First, high angular resolution diffusion imaging (HARDI) techniques are used to estimate the fiber orientation distribution function (ODF) at each location \citep{Descoteaux2009} (implemented in {\em dipy} \citep{Garyfallidis2014}). Next, streamlines following the principle directions of the fiber ODF are  constructed by probabilistic tractography algorithms under local continuity constraints.   Anatomical structure information is used to guide selection of where to start and stop the streamlines \citep{Smith2012,Girard2014266}.   The final constructed 3D curves are assumed to represent the most likely pathways through the diffusion profile delineated by the fiber ODF.  We refer to these curves as fibers, though they may not exactly correspond to anatomical fibers in the brain.

\begin{figure}
\begin{center}
\includegraphics[height=2.4in]{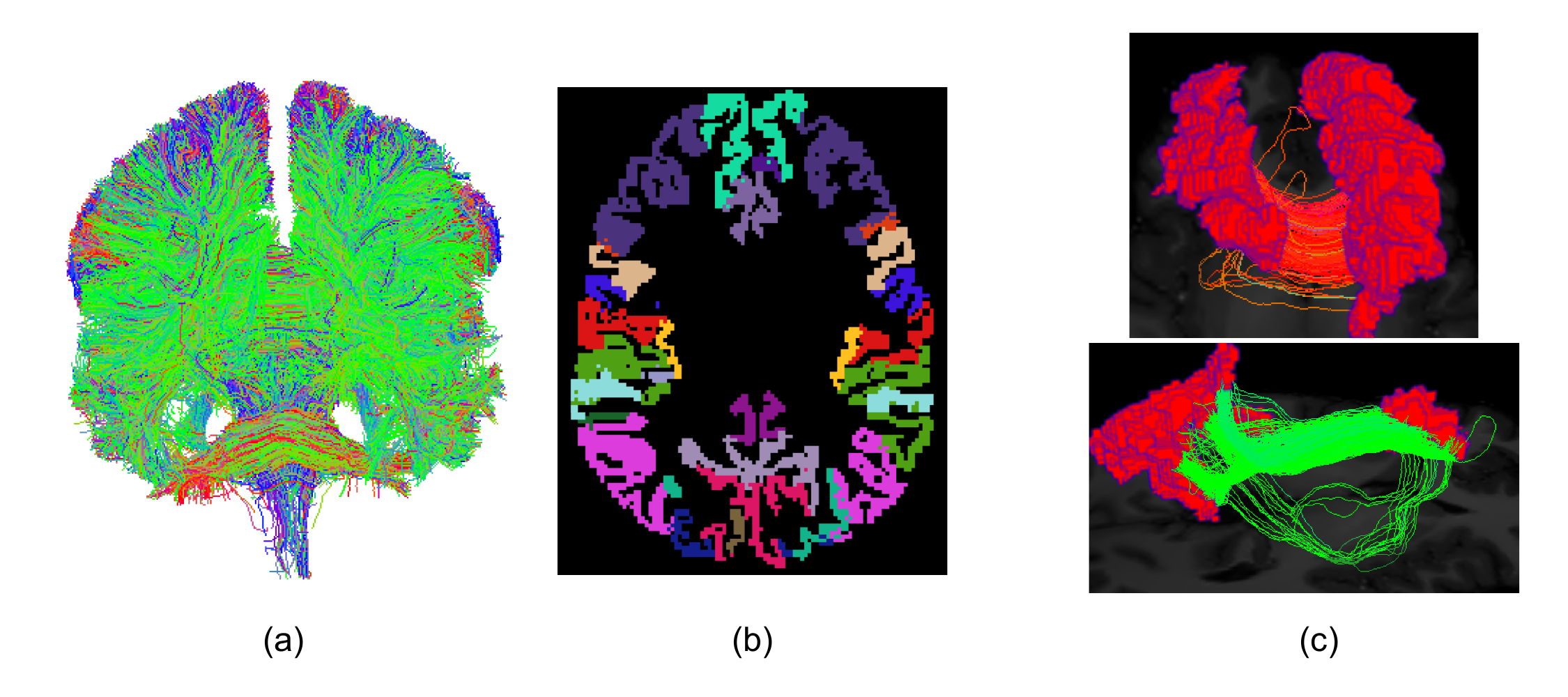} 
\caption{(a) shows one example of the whole tractography dataset for a brain; (b) shows the Desikan-Killiany parcellation of the cortical region and (c) shows fiber curves connecting a pair of regions in the Desikan-Killiany parcellation. } \label{fig:pipe_ill2}
\end{center}
\end{figure}

Let $T_j$ denote the $j$th subject's tractography dataset. In general, $T_j$ contains millions of fiber curves indicating how different regions of the brain are connected. Let $y_{ji}$ represent a single fiber curve in $T_j$; the data on $y_{ji}$ output by the tractography algorithm consist of hundreds of points along a curve, but we view $y_{ji}: [0,1] \rightarrow \Re^3$ as a parameterized curve that can be accurately approximated by spline interpolation of these data points.  Figure \ref{fig:pipe_ill2} (a) shows one example of the tractography dataset we generated for an individual's brain. 

Directly analyzing all fibers in $T_j$ is not realistic for several reasons. The data are huge (millions of fibers in each subject) and current statistical methods are ineffective in handling such big data for a sample of subjects. Secondly, the streamline datasets are usually in subject-specific spaces with different coordinate systems, and it is hard to directly compare any two tractography datasets.  Alignment between different subjects is necessary to define a realistic probability model for multiple subjects, but there are currently no effective tractography alignment methods.

In this paper, we group each fiber in $T_j$ based on the different anatomical regions it connects, and focus our analysis on fibers connecting two specific regions of interest.  To achieve this, each individual's brain is first parcellated into different meaningful anatomical regions based on an existing template, such as the Desikan-Killiany atlas \citep{Desikan2006968}. Figure \ref{fig:pipe_ill2} (b) shows the parcellation of the brain using the Desikan-Killiany atlas. Then fiber curves connecting each pair of regions are extracted, as illustrated in Figure \ref{fig:pipe_ill2} (c).

Our goal is to build a flexible but parsimonious Bayesian model to characterize the distribution of fiber curves connecting two regions $r_a$ and $r_b$ within each individual and across a population of individuals. The extracted fibers $\{ y_{ji}, i=1,...,n_j\}$ connecting $r_a$ and $r_b$ in subject $j$ have some special properties, e.g. $y_{ji}$'s always start from one region and end at another one and they are smooth and follow similar white matter pathways.  These properties make the underlying functional space ${\Omega}_{(r_a,r_b)} $ much smaller comparing with $\Omega$, where $\Omega$ is the entire functional space $\mathcal{L}([0,1],\Re^3)$ and ${\Omega}_{(r_a,r_b)} $ is the functional space for fiber curves connecting $r_a$ and $r_b$ for all subjects in our dataset.  If one fits statistical models with a support of $\Omega$,  these models will tend to be inefficient at capturing the data. In order to build an efficient model on the correct space parsimoniously, we consider a variance decomposition for fibers in $\Omega_{(r_a,r_b)}$.


\subsection{Variation decomposition}

When we treat a fiber $y$ as a 3D curve, there are five factors contributing to the variance: (1) translation, (2) rotation, (3) scaling, (4) re-parameterization and (5) shape. Translation, rotation, scaling and re-parameterization are shape-preserving transformations \citep{anuj2011}.   The shape of a fiber represents appearance after removing these shape-preserving transformations. Letting $y \in \mathcal{L}([0,1],\Re^3)$ be a fiber, a translation of $y$ is represented as $y+a$, where $a \in \real^3$. The rotation of $y$ is represented as $\bO*y$, where $\bO \in SO(3)$ is a rotation matrix. Scaling represents the length of the fiber.  Re-parameterization of $f$ is represented as $y(\gamma(s)), s\in[0,1]$, where $\gamma$ is a warping function in $\Gamma$, the set of all orientation-preserving diffeomorphisms of $[0,1]$. Note that re-parameterization of $f$ does not change the shape, it only changes the point-wise correspondence between fibers. In other words, if we let $g(s) = y(\gamma(s))$, $g$ passes through the same path as $y$, but $g(s)$ is different from $y(s)$ if $\gamma(s) \neq s$.  The re-parameterization component performs the role of aligning different fibers \citep{Tucker:2013,kurtek2012}, and this alignment does not change the path of each fiber, but reduces the shape component of variability.

\begin{figure}
\begin{center}
\begin{tabular}{cccc}

\includegraphics[height=1.1in]{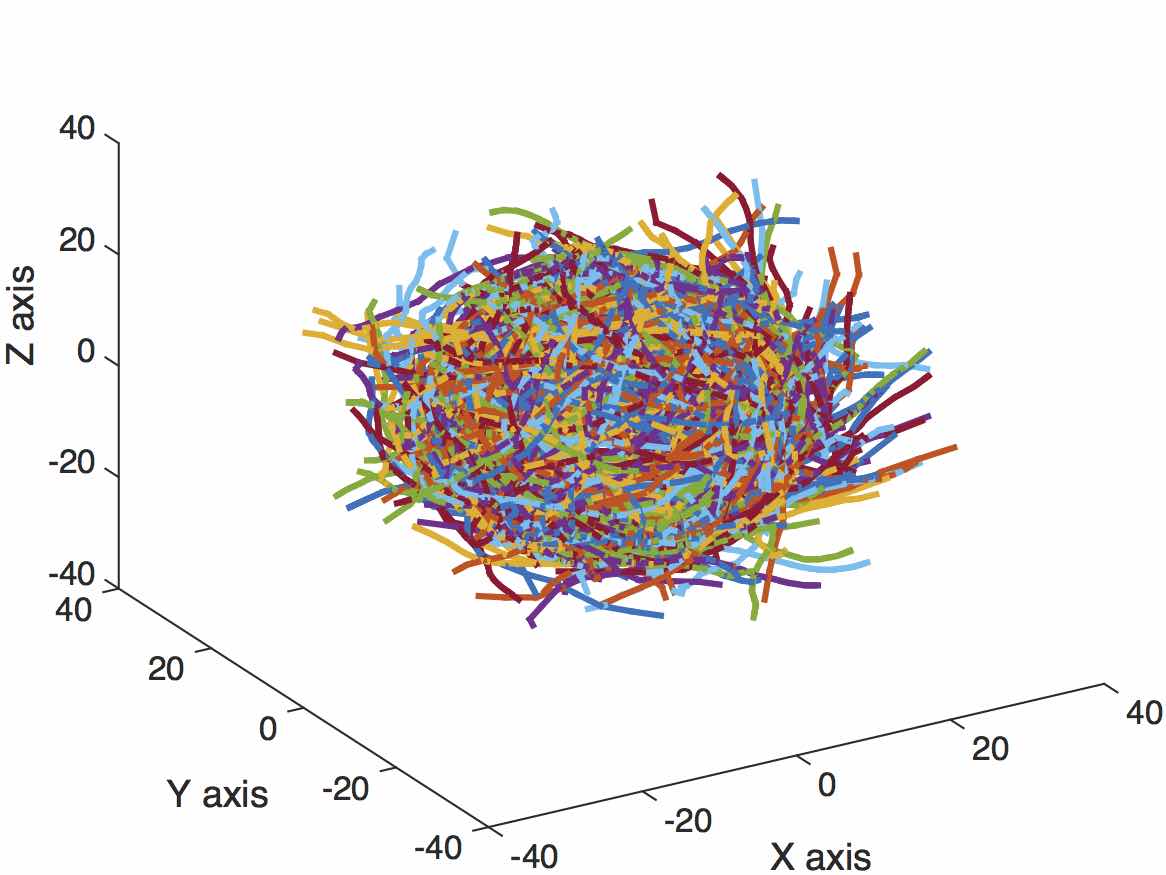} &
\includegraphics[height=1.1in]{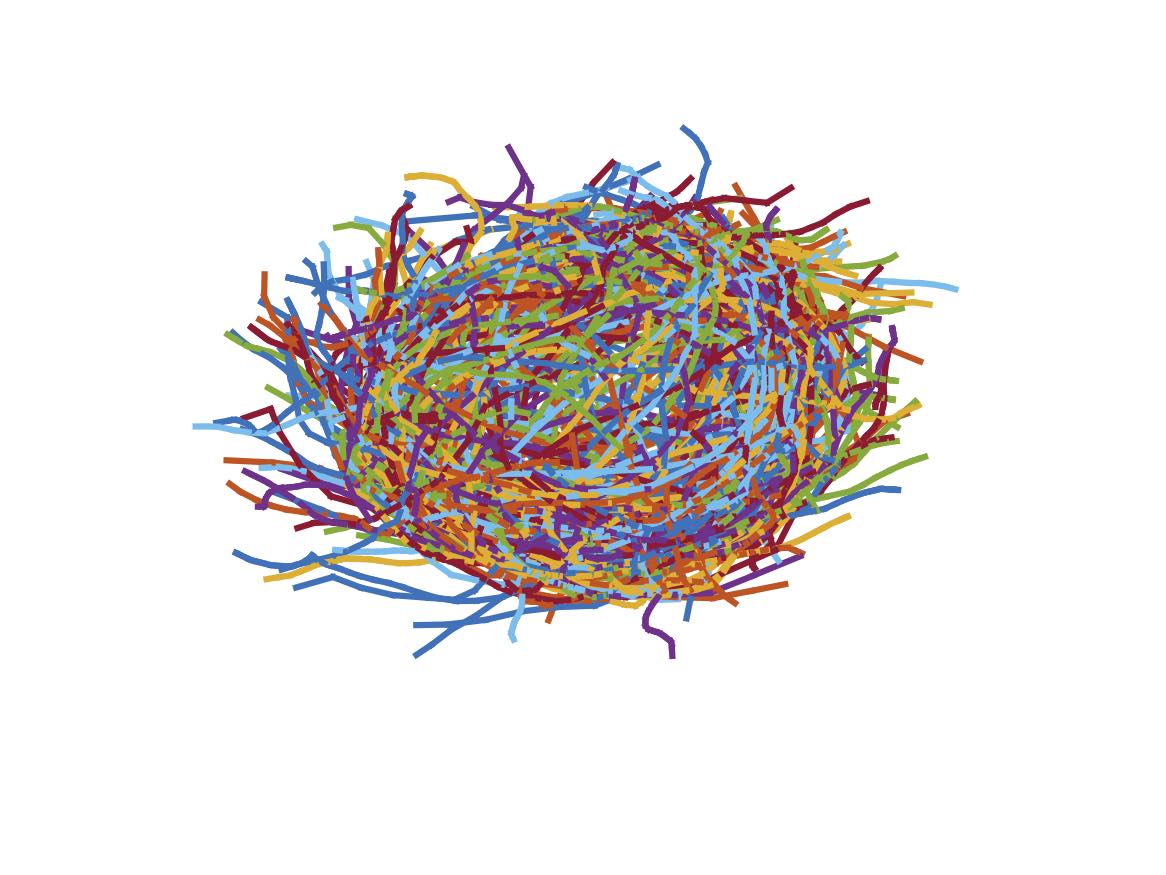}&
\includegraphics[height=1.1in]{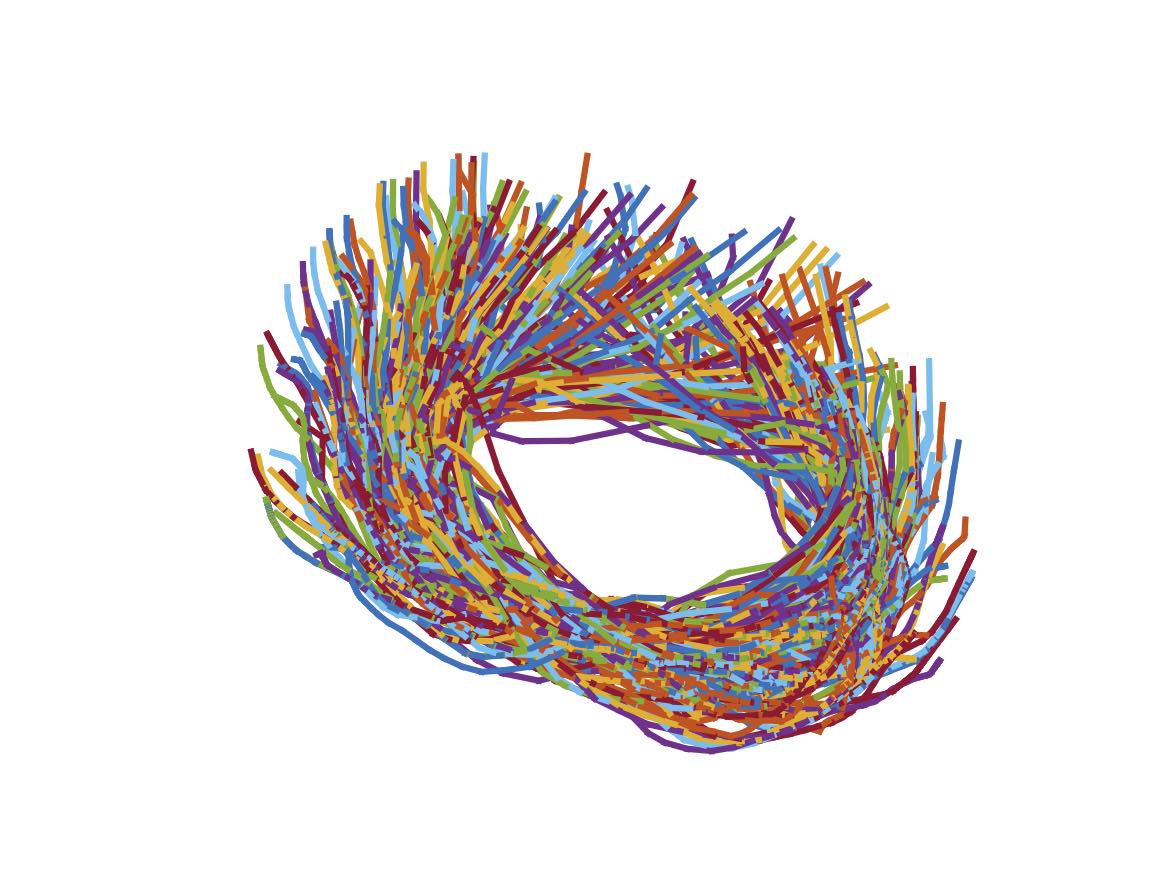}&
\includegraphics[height=1.1in]{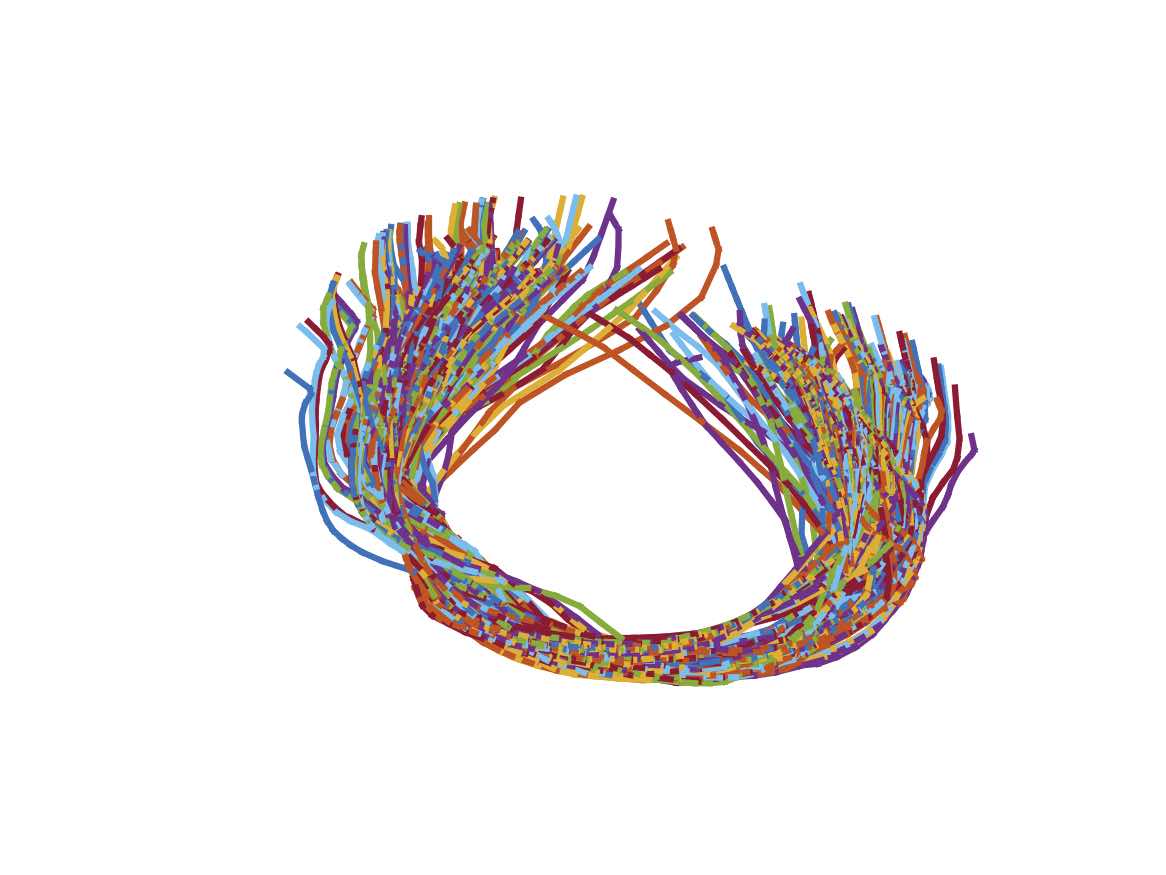}\\
\includegraphics[height=1.0in]{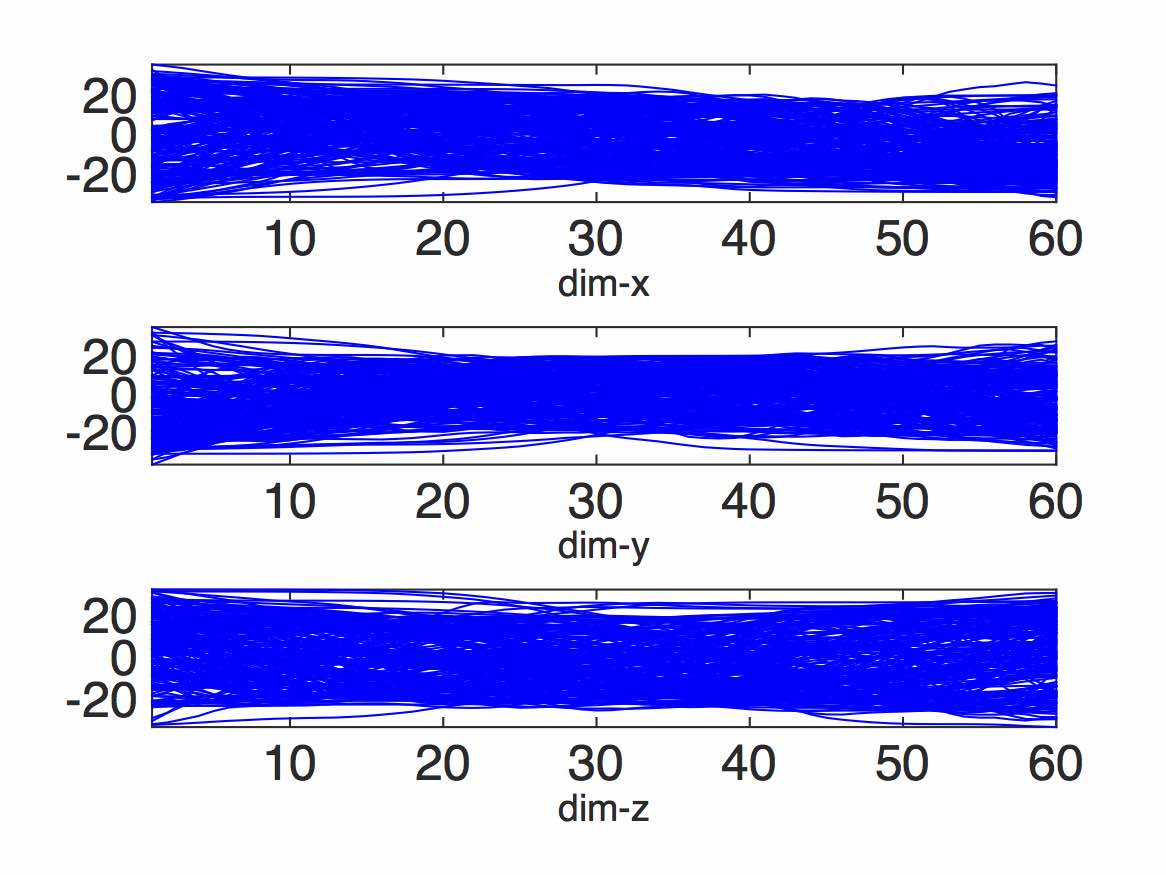}&
\includegraphics[height=1.0in]{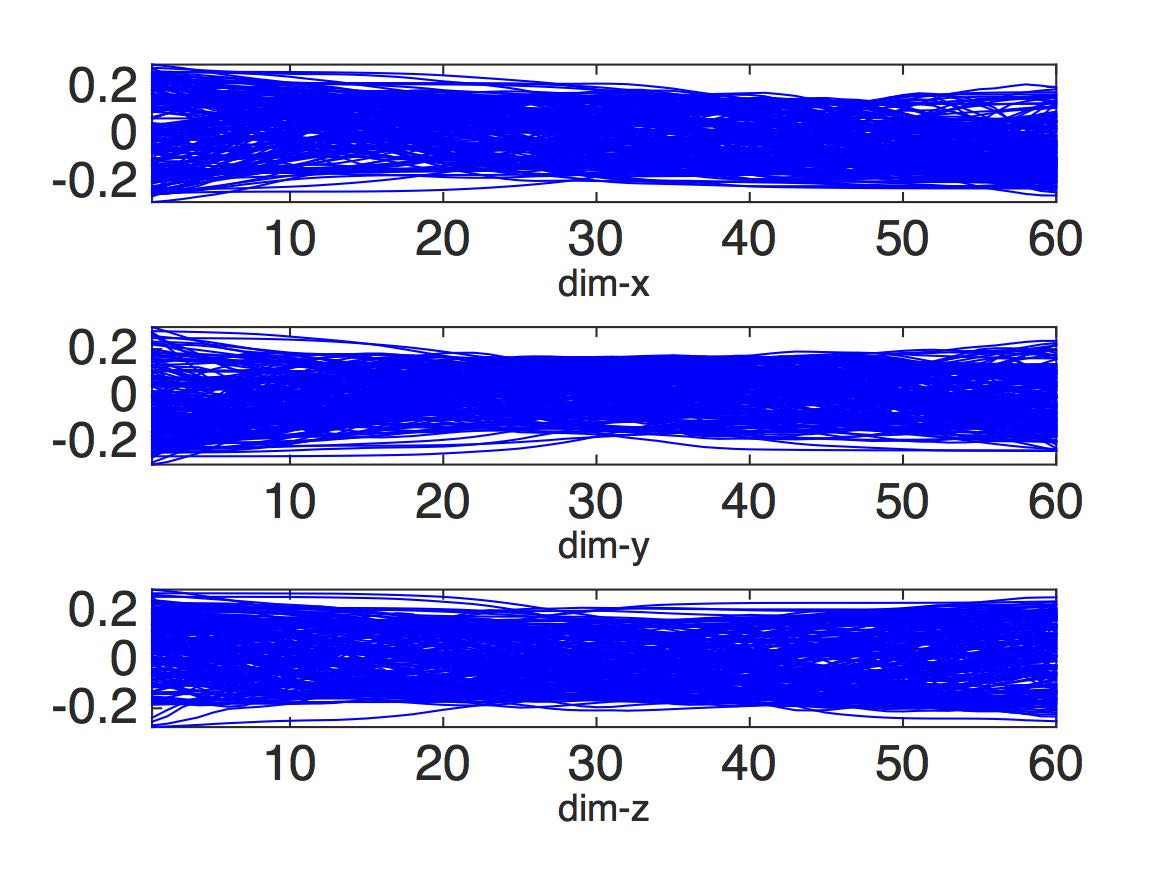}&
\includegraphics[height=1.0in]{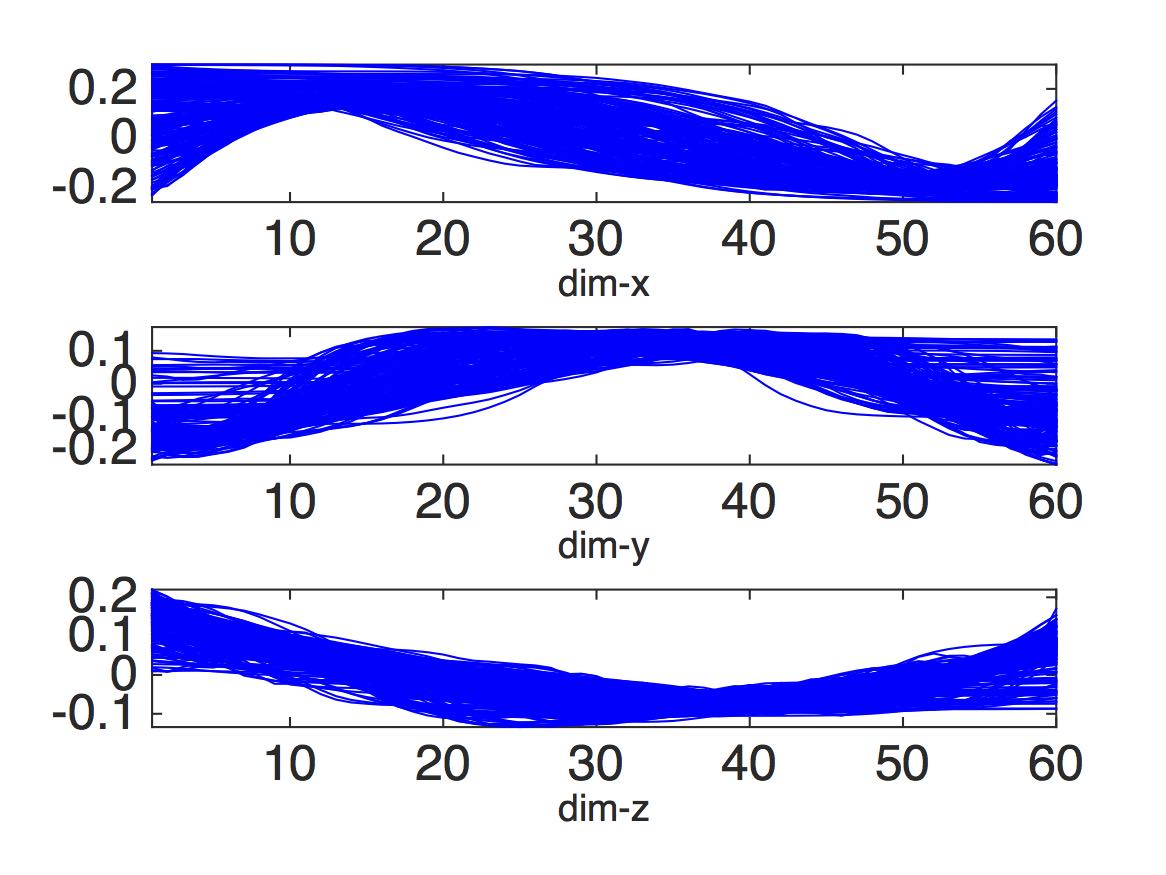}&
\includegraphics[height=1.0in]{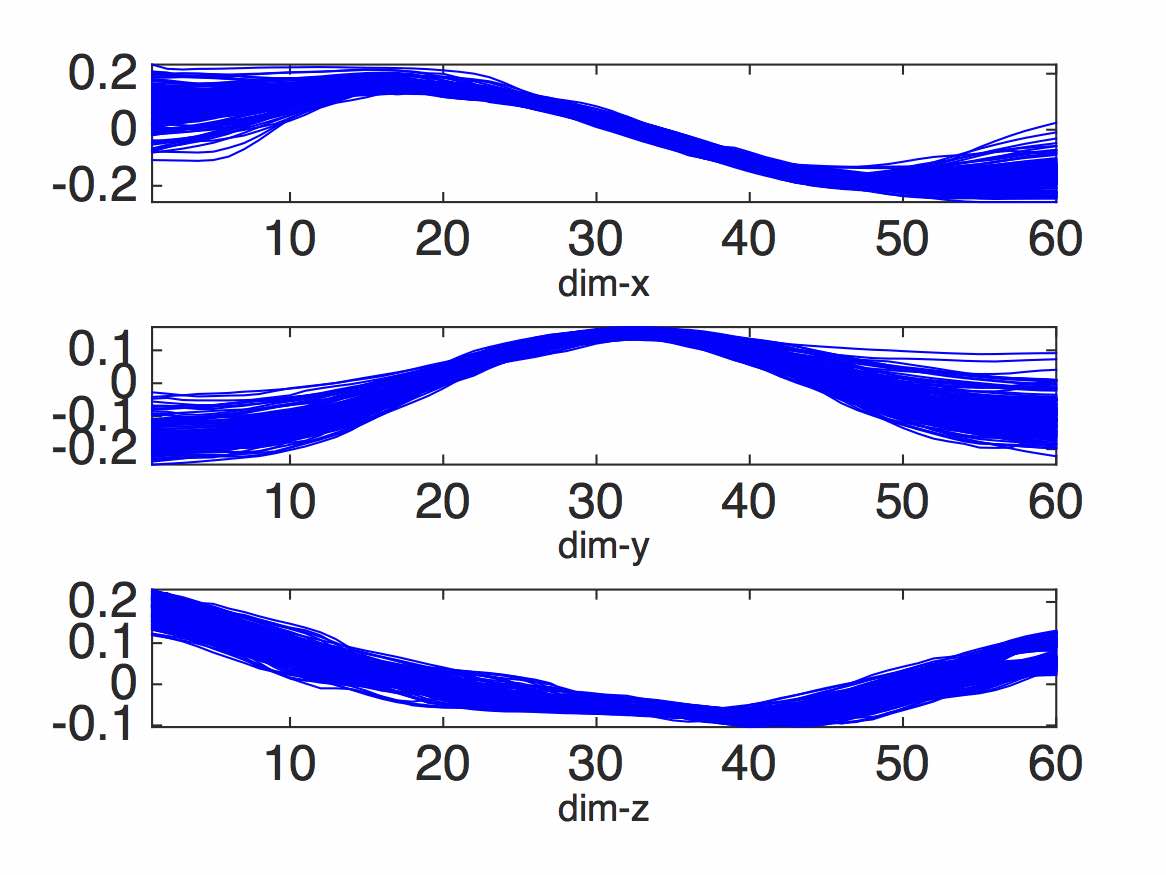} \\
(a) & (b) & (c) & (d) \\
\end{tabular}
\caption{The remaining shape components after removing different shape-preserving transformations. (a) shows the simulated raw 3D curves. (b) shows the shape part after removing translations and scalings. (c) shows the shape part after removing translations, rotations and scalings. (d) shows shape part after removing translations, rotations, scalings and re-parameterizations.  } \label{fig:fiber_sim1}
\end{center}
\end{figure}

Figure \ref{fig:fiber_sim1} illustrates the shape components after removing different shape-preserving transformations for $200$ simulated fiber curves. As additional shape-preserving components are separated out, the remaining shape part has decreasing  cross-sectional variance at each point $s \in [0,1]$.  Since in our particular case, fibers connecting two regions of interest usually have similar length, we do not remove scaling.  In addition, the re-parameterization component does not contribute to the geometric appearance of fibers, i.e. the path of the fibers, and thus we treat it as a confounding variable, which is removed in an alignment phase prior to statistical analysis.  

The main motivation for us to perform this decomposition is that the variance in each component after decomposition becomes much smaller, allowing us to more effectively model each component separately while inducing a flexible joint model. This is especially true for the shape part. After separating the shape-preserving transformations, the remaining shape part of the fibers are aligned together.  This alignment can be done within and between subjects, which means that we align all fibers from different subjects together. This procedure naturally solves the misalignment issue for analyzing fibers in a set of subjects.  In addition, since we only consider fibers connecting the same anatomical regions, we find that these fibers have similar shapes, as illustrated in Figure \ref{fig:fiber_ill1}. Therefore, the variation decomposition process enables us to learn a low dimensional structure for the shape part of the fibers in each connection for all subjects. The other parts, such as the translation and rotation, can be efficiently represent with very few parameters.  This process produces a low dimensional representation for fiber curves in a connection.  

\subsection{Estimating manifold components from curves}
As a preliminary step before defining a Bayesian model, we extract each component in the variance decomposition by using the elastic shape analysis framework of \cite{anuj2011}. Given a set of fiber curves $\{y_1,...,y_n \}$ in one connection,
to separate the translation, we  center  each fiber by ${y}_i^c(\cdot) = {y_i(\cdot)-c_i^{(1)}}$, where $c_i^1 = \int_0^1 y_i(s) |\dot{y}_i(s)|ds/L_{y_i}$, in which 
$\dot y_i(s)=dy_i(s)/ds$ and $L_y$ is the length of fiber $y$. Without special notation, all fibers have been centered from now on. To separate the rotation and re-parameterization, we represent each fiber $y$ as its square-root velocity function (SRVF) $q(s)$, defined as $q(s) = \dot{y}(s)/\sqrt{\|\dot{y}(s)\|}$. A rotation of $y$ by $\bO \in SO(3)$ is denoted as $\bO*y$  and its SRVF becomes $\bO*q$.  A re-parameterization of $y$  by $\gamma \in \Gamma$ is denoted as $y(\gamma(s))$, and its SRVF is denoted as $(q,\gamma)=(q\circ\gamma)\sqrt{\dot{\gamma}}$, where $\circ$ denotes the composition of two functions.  The motivation for using SRVF is that it allows us to use a well-known elastic metric, the Fisher-Rao Riemannian metric, to perform elastic shape analysis, i.e. separating the re-parameterization from the shape part of the fibers.

To align all fibers by separating the rotation and re-parameterization, one needs to estimate a template fiber first, denoted as $y_\mu$, and then align all fibers to the template. We formulate the calculation of $y_\mu$ and individual alignment as an iterative procedure: first  initialize the mean function $y_\mu$ and its SRVF $q_{\mu}$ and then iteratively solve for 
\begin{equation} 
(\bO_i,  \gamma_i) =  \argmin_{\bO \in SO(3), \gamma \in \Gamma} \|q_{\mu}-\bO*(q_i, \gamma) \|~\mbox{, and}~
q_{\mu}= n^{-1}\sum_{i=1}^{n} \bO_i*(q_i, \gamma_i)   
\end{equation}
for $ i=1, \ldots, n$ until convergence.   
We optimize   $\bO_i$   through Procrustes analysis and   $\gamma_i$  through dynamic programming \citep{anuj2011}. As the output of this iterative algorithm, for each fiber $y_i$, we obtain the best rotation $\bO_i$, re-parameterization $\gamma_i$, to the template $y_\mu$, and the shape part $g(s) = \bO_i*y_i(\gamma_i(s))$. 

To efficiently represent the shape part of the fibers in the connection of $(r_a,r_b)$, we use FPCA to learn basis functions representing the aligned fiber curves.  We learn an FPCA basis $\{\phi_{l}:[0,1] \rightarrow \Re^3, l =1,...,T\}$ using training data from healthy subjects.  FPCA can characterize the variations within the given
training data set and extract the principal modes of deformations of the fibers relative to
the mean fiber. For each pair of ROIs, these basis functions only need to be learnt once and can be saved for further use.
For the connection $(r_a,r_b)$, we obtain a low-dimensional structure consisting of
$$ L_{(r_a,r_b)} = \{y_\mu,\{\phi_{l}, l=1,...,T \}  \}. $$
Letting $g$ be the shape part of fiber $y \in \Omega_{(r_a,r_b)}$, we can represent $g$ as 
$$g(s) \approx y_{\mu}(s) + \sum_{l=1}^T x_l \phi_{l}(s), $$
where $x_l$ represents the coefficient corresponding to $\phi_l$. For notational convenience, we let $c^{(2)} = [x_1,x_2,...,x_T]' \in \Re^T$.   

We decompose fiber curve $y \in \Omega_{(r_a,r_b)}$ as $y(s):=\{c^{(1)},c^{(2)},\bO, \gamma\}$, where $c^{(1)}, c^{(2)}, \bO$ and $\gamma$ are the translation, shape, rotation and reparameterization components, respectively. The original fiber path can be recovered from these components using 
$$\hat{y}(s) \approx \bO^T* \big( y_\mu + \sum_{l=1}^T c^{(2)}(l)\phi_l \big) +c^{(1)}.  $$
The difference between the recovered path $\hat{y}$ and the original path $y$ depends on the representation precision of the shape part: with more basis functions, one can more precisely recover the shape part and thus the original fiber path.  We did not include $\gamma$ in the recovery formula because $\gamma$ does not change the geometric appearance of $y$ but only changes its parameterization. In this paper, we focus on modeling the geometry of the fiber curves, and therefore, $\gamma$ is excluded.

\section{ Model for one individual} \label{sec:individual}
\subsection{Product kernel mixture model}
In this section, we model fiber curves from a single subject. Let $\{ y_i\}$ for $i=1,...,n$ be the fiber curves connecting a pair of regions of interest in an individual. After the decomposition, each fiber $y_{i}$ is represented as $y_{i} := \{ c^{(1)}_{i}, c^{(2)}_{i},\bO_{i}\}$. For notation convenience, we denote $c^{(3)}_i = \bO_{i}$, and we have $y_{i} := \{ c^{(1)}_{i}, c^{(2)}_{i},c^{(3)}_{i}\}$. Each of the $c^{(m)}_{i}$ (for $m=1,..,M$) has a different Euclidean or manifold support. Letting $c^{(m)}_{i} \in \mathcal{Y}_m$,  we have 
$$y_{i} \in \mathcal{Y} = \bigotimes _{m \in I}\mathcal{Y}_m , I = \{1,...,M\}. $$ 
Our goal is to specify a joint model in which $y_i \sim f$, with $f$ a probability measure characterizing the joint distribution.  Let $\mathcal{B}(\mathcal{Y})$ denote an appropriate $\sigma$-algebra of $\mathcal{Y}$, with $f$ assigning probability $f(B)$ to each $B \in \mathcal{B}(\mathcal{Y} )$. 
 
Initially, we focus on modeling one component of $y_i$, the $m$th component $c^{(m)}_i$. A straightforward strategy is to use a mixture model with 
\begin{equation} \label{eqn:ind}
f_m(c) = \int_{\Theta_m} \mathcal{K}_m(c;\theta^{(m)})dP(\theta^{(m)}),\quad c \in  \mathcal{B}(\mathcal{Y}_m),
\end{equation}
where $\mathcal{K}_m(\cdot;\theta^{(m)})$ is a parametric probability measure on $\{\Y_m,\mathcal{B}(\Y_m)\}$, and $P$ is a probability measure over $\{ \Theta_m, \mathcal{B}(\Theta_m) \}$. A nonparametric Bayesian approach is realized by choosing $P$ as a random probability measure and assigning an appropriate prior through 
\begin{equation} \label{eqn:sigprior}
P=\sum_{h=1}^K \pi_h \delta_{\theta_h},\quad \theta_h \sim P_0^m,
\end{equation}
where $P_0^m$ is a base measure on $\{ \Theta_m, \mathcal{B}(\Theta_m) \}$ and $\delta_{\theta_h}$ denotes a degenerate distribution with all its mass at $\theta_h$. Equation (\ref{eqn:sigprior}) contains a broad class of species sampling priors, including, for example, the Dirichlet process and Poisson-Dirichlet process. In the Dirichlet process case, $K = \infty$ and $\pi_h$ is generated through a stick-breaking process \citep{Sethuraman1994}, with $\pi_h = V_h \prod_{l<h}(1-V_l)$ and $V_h \sim \text{beta}(1,\alpha)$ independently for $h=1,...,\infty$.

To jointly model the different components of $y_i$, we apply a product kernel mixture as in \citep{Bhattacharya2012b,BanerjeeMD13}.  In particular, supposing that 
$y_i \stackrel{iid}{\sim} f$, 
\begin{equation} \label{likelihood:y}
f(y_{i}) =  \int_{\Theta} \prod_{m=1}^M \mathcal{K}_m(c_{i}^{(m)};\theta^{(m)})dP(\theta),\quad \theta=\{\theta^{(1)},...,\theta^{(M)} \} \  ,
\end{equation}
where $\mathcal{K}_m$ is a parametric density on $\mathcal{Y}_m$, and $P$ is a mixing measure with the form,
\begin{equation} \label{mixture:y}
 P = \sum_{h=1}^K \pi_h \delta_{\theta_h},\quad \theta_h = \{\theta_{h}^{(1)},\theta_{h}^{(2)},\cdots, \theta_{h}^{(M)}\}\sim P_0 = \prod_{m=1}^M P_{0}^m. 
 \end{equation}

Under this model, the conditional likelihood for fiber $y := \{ c^{(1)},\ldots,c^{(M)} \}$ given $\pi = \{\pi_1,...,\pi_k\}$ and $\theta$ can be written as
\begin{equation} 
f(y|\pi, \theta) = \sum_{h=1}^K \pi_h \prod_{m=1}^3 \mathcal{K}_m(c^{(m)};\theta_h^{(m)}).
\end{equation}
  Introducing a cluster index $S_i \in \{1,\ldots,k\}$ for fiber $i$,  we have
$c_i^{(m)} \sim \mathcal{K}_m(\cdot;\theta_{S_i}^{(m)})$ independently for $m=1,\ldots,M$, and $\mbox{Pr}( S_i = h ) = \pi_h$, for $h=1,\ldots,K$.  This conditional 
independence structure given the cluster indices of the fibers facilitates computation, while still allowing a flexible dependence structure between the 
different components marginally.  The remaining task is to specify the $\K_m(c_i^{(m)};\theta^{(m)})$ for each component. 

\subsection{Kernel density for each component} 
We describe the intrinsic space of each component $m$ and define a parametric distribution $\mathcal{K}_m$ having appropriate support. We have $M=3$ corresponding to the translation ($m=1$), rotation ($m=2$) and shape ($m=3$) components.

{\bf Translation Component:} The translation component $c^{(1)}$ is a vector in $\Re^3$. We simply use a multivariate normal distribution for $c^{(1)}$,

$$\mathcal {K}_1 (c^{(1)} ; \mu , {\bf \Sigma}) = \frac{1}{\sqrt{(2\pi)^3| {\bf \Sigma} |} }  \exp\Big\{ - {1\over 2} {{(c^{(1)} - \mu)^T {\bf \Sigma}}^{-1} (c^{(1)} - \mu) } \Big\}.
$$

{\bf FPCA Coefficient Component:} Let $c^{(2)} \in \Re^T$ denote the shape component corresponding to the coefficients of the FPCA basis functions. Similar to the translation component, we assign a multivariate normal distribution for $c^{(2)}$,
$$\mathcal{K}_2 ({c}^{(2)}; \mu, {\bf \Sigma})  =  \frac{1}{\sqrt{(2\pi)^{T} | {\bf \Sigma} |} }  \exp\Big\{ - {1\over 2} {{(c^{(2)} - {\bf \mu})^T {\bf \Sigma}}^{-1} (c^{(2)} - {\bf \mu}) } \Big\}. $$

{\bf The Rotation Component}:  The rotation matrix ${ c^{(3)}}$ is an element of the special orthogonal group $SO(3) = \{ { {\bf X} \in} O(3)| \det({\bf X}) = 1  \}$.  The most common parametric 
distribution on $SO(3)$ is the matrix Fisher distribution, also known as the Langevin distribution \citep{Downs1972,Khatri1977,Jupp1979}.  \citet{Bingham2009} and \citet{Qiu2014}
proposed a more flexible class of Uniform Axis Random Spin (UARS) distributions, which improves upon the flexibility of the Langevin.  We carefully considered both choices, but faced 
computational and stability problems in conducting inferences, particularly as the number of fibers increases.

To address these problems and take advantage of the similarity of the decomposed rotation matrices to the identity, we define a simple Gaussian like parametric distribution based on an embedding in the Lie algebra of $SO(3)$. Let ${ {\bf I}_3}$ denote the identity element of $SO(3)$. The tangent space at ${ {\bf I}_3}$, $T_{ {\bf I}_3}{(SO(3))}$ forms a Lie algebra, which is usually denoted as $\mathfrak{so}(3)$. The exponential map, $\exp: \mathfrak{so}(3) \rightarrow SO(3)$, provides a mapping from the tangent space $T_{ {\bf I}_3}({SO(3)})$ to $SO(3)$. The inverse of the exponential map is called the $\log$ map. $\mathfrak{so}(3)$ is a set of $3\times 3$ skew-symmetric matrices. 
 We use the following notation to denote any matrix ${{{\bf A}_v} }\in\mathfrak{so}(3) $:
\begin{equation} \nonumber
{{\bf A}_{v}} = 
 \begin{pmatrix}
  0 & -v_1 & v_2  \\
  v_1 & 0& -v_3  \\
  -v_2 & v_3 & 0
 \end{pmatrix}
 \end{equation}
 where ${v}=[v_1,v_2,v_3] \in \Re^3$.  The exponential map for $\mathfrak{so}(3)$ is given by Rodrigues' formula, 
 \[ \exp({ \bf A_v}) =
  \begin{cases}
    { \bf I}_3,      & \quad \alpha = 0\\
    {\bf I}_3 + \frac{\sin(\alpha)}{\alpha} { {\bf A}_v} + \frac{1-cos(\alpha)} {\alpha^2}  { {\bf A}_v} ^2,    & \alpha \in (0,\pi) , \\
  \end{cases}
\]
where $\alpha = \sqrt{{1\over 2}  tr ({ {\bf A}_v}^T{{\bf A}_v } )} = \| v\| \in [0,\pi)$. The $\log$ map for a matrix ${ \bf X} \in SO(3)$ is a matrix in $\mathfrak{so}(3)$, given by 
 \[ \log({ \bf X}) =
  \begin{cases}
    {\bf 0},      & \quad \alpha = 0\\
    {\alpha \over 2 \sin(\alpha)}({ \bf X}-{\bf X}^T,)    & |\alpha| \in (0,\pi) , \\
  \end{cases}
\]
where $\alpha$ satisfies $tr({ \bf X})=2\cos(\alpha) + 1$.

Define a mapping $\Phi$ to embed an element in $\mathfrak{so}(3)$ to $\Re^3$, $\Phi: \mathfrak{so}(3) \rightarrow \Re^3$, $\phi({ \bf A}_v) = { v}$. Let ${ \phi (\bf X)} = \Phi(\log({ \bf X})) \in \Re^3$ be the embedded vector for the element $ { \bf X}\in SO(3)$ in $\real^3$.  We define a trivariate normal distribution on this embedding space:
\[
\mathcal{K}_3({\bf X}; \mu,{\bf  \Sigma}) = \frac{1}{\sqrt{(2\pi)^3| {\bf \Sigma} |} }  \exp\bigg\{ - {1\over 2} {({\phi({\bf X})-\mu) {\bf \Sigma}}^{-1} ({\phi({\bf X})-\mu)}^T} \bigg\},
\]
where $v$ is an element in the embedding space $\Re^3$.  This embedded Gaussian kernel has substantial practical advantages over alternative intrinsic parametric kernels we attempted to implement.

\subsection{Prior specification and posterior inference} \label{sec:priorpost}

To complete a Bayesian specification of the model, we choose a prior for the cluster probabilities: 
$$\pi = (\pi_1,\ldots,\pi_K)' \sim Dir\bigg( \frac{\alpha}{K},\ldots, \frac{\alpha}{K} \bigg),$$
where $K$ is an upper bound on the number of clusters.  In the limit as $K \to \infty$, this choice leads to a Dirichlet process mixture model.  In addition, 
\cite{Rousseau2011} 
motivated a similar choice of prior as being effective as favoring deletion of redundant mixture components not needed to characterize the data. If $K$ is chosen to be too small, then none of the clusters will be unoccupied, and the analysis should be repeated for larger $K$.  

Posterior sampling proceeds via the following steps:
\begin{enumerate}
\item Update the cluster allocation of $S_i$ for each fiber curve from the conditional posterior with 

$$\text{Pr}(S_i = h| -) =  {\pi_h \prod_{m=1}^M \mathcal{K}_m(c_i^{(m)};\theta_h^{(m)}) \over \sum_{l=1}^K \pi_l \prod_{m=1}^M \mathcal{K}_m(c_i^{(m)};\theta_l^{(m)})}. $$

\item Update the weights on each component from the conjugate conditional posterior
$$(\pi| - ) \sim Dir\bigg(\frac{\alpha}{K}+n_1,...,\frac{\alpha}{K}+n_K\bigg),$$
where $n_h$ is the number of observations with cluster $h$.

\item Update the parameter $\theta_h^{(m)}$ for $m=1,...,M$ and $h = 1,...,k$ from 
$$(\theta_h^{(m)}|-) \propto  P_0^m(\theta_h^{(m)}) \prod_{i:S_i=h} \mathcal{K}_m(c_i^{(m)};\theta_h^{(m)}),$$
\end{enumerate}
where $P_0^m$ is a conjugate prior to $\mathcal{K}_m(c_i^{(m)};\theta_h^{(m)})$ for each component $m$; in particular, 
we use Gaussian-Inverse Wishart priors.

\section{ Model for a population of individuals}
Section \ref{sec:individual} proposes a flexible mixture model for the distribution of fibers connecting a pair of regions of interest in a single individual's brain; in this section, we generalize the model to accommodate multiple individuals. This generalization is challenging because (1) fibers in each individual have their own coordinate system inherited from the diffusion MRI scan; (2) there are different numbers and appearances of fiber curves for different individuals.  Although (1) can potentially be addressed via image alignment before or during tractography, such alignment is not straightforward. Our variation decomposition bypasses  this issue by building a common coordinate system for the fiber shapes.  Issue (2) can be solved by using a hierarchical Bayesian model to allow differences between individuals while encouraging borrowing of information.  

  Let $\{y_{ji}\}$ for $j=1,..,J$ and $i = 1,...,n_j$ be a collection of fiber curves for the same pair of brain regions in $n$ subjects, where $n_j$ represents the number of fiber curves in the $j$th subject. We have $y_{ji} =\{c_{ji}^{(1)},...,c_{ji}^{(M)}\}$, so that the fibers are represented by their different geometric components.  In addition, let $w_j$ denote a scalar summary of the strength of connection between the brain regions for individual $j$. In the literature, $w_j$ is usually set as the number of fibers, $n_j$. 

\subsection{Nested Dirichlet process model} \label{sec:fc}

The model in Section \ref{sec:individual} allows the distribution $f$ of fiber curves within an individual to be unknown.  Generalizing to multiple individuals, we have distributions 
$f_j$, for $j=1,\ldots,J$, and require a model for an unknown distribution of distributions, $f_j \sim Q$, with $Q$ unknown.  One natural possibility is a hierarchical Dirichlet process (HDP) mixture \citep{Yee2015}, which would induce clusters of fibers, with these clusters having different weights for each individual $j$.  This type of model would effectively assume that white matter pathways (each pathway represents a cluster) connecting two regions of interest are shared by all individuals, but the proportions of fiber curves in each pathway are different.  However, we found that this type of model has poor performance, as our data (illustrated in Figure 1) show that many subjects have completely different white matter fiber bundles.  This motivates us to instead use the nested Dirichlet process (NDP) \citep{Abel2008}, which clusters subjects based on their fiber curve distribution, with subjects in a cluster having similar clusters of fibers.  

Our NDP model has the following form: 
 \begin{eqnarray} \label{eqn:ndp}
f_j(y_{ji}) &=& \int \prod_{m=1}^M \mathcal{K}_m(c_{ji}^{(m)};\theta^{(m)})dG_j(\theta),\quad \theta = \{\theta^{(1)},...,\theta^{(M)}\}, \nonumber  \\
G_j(\cdot) & \sim & \sum_{h=1}^\infty \pi_h^* \delta_{G^*_h(\cdot)},\quad G^*_h(\cdot) = \sum_{l=1}^\infty \omega_{lh}^* \delta_{\theta_{lh}^*}, \label{eqn:ndp1}
 \end{eqnarray}
where $\theta^*_{lh} = \{ \theta_{lh}^{(1)*},...,\theta_{lh}^{(M)*} \}$ and $\theta^*_{lh} \sim \prod_{m=1}^M P_0^m$, $\omega_{lh}^* = u_{lh}^*\prod_{s=1}^{l-1}(1-u_{sh}^*)$, $\pi^*_h=v_h^* \prod_{s=1}^{h-1}(1-v_s^*)$, $v_h^* \sim \text{ beta}(1,\alpha)$, and $u_{lh}^* \sim \text{ beta}(1,\beta)$. The collection of individual-specific mixing measures $\{ G_j \}$ are drawn 
from an NDP, $\{ G_j \} \sim \text{NDP}(\alpha, \beta, P_0)$, where $P_0 = \prod_{m=1}^M P_0^m$ is the base measure.  

Under this structure, the prior probability that two individuals are assigned to the same brain structure cluster is $1/(1+\alpha)$, while the prior probability of clustering two fibers together within a brain is $1/(1+\beta)$.  The model can be used for any combination of the components of variability in the fiber curves; for example, 
 one can use only the shape component or a combination of different components to estimate $f_j$.  In applying these models to brain connectomics data, we will assess how clustering performance depends on which components are included.

\subsection{Posterior inference} \label{sec:pp_ps}

Following \cite{Abel2008}, we propose a blocked Gibbs sampling algorithm. An approximation of the stick-breaking process is used, with the infinite sums in (\ref{eqn:ndp}) replaced by
finite sums of $K$ (for $G_j$) and $L$ (for $G^*_h$) elements.
Let $\zeta_j$, for $j=1,...,J$, be the membership indicator of individuals and let $\xi_{ji}$, for $i=1,...,n_j$, be the membership indicator of fiber curves for the $j$th subject. Sampling proceeds via the following steps:
\begin{enumerate}
\item Sample the membership indicator for the $j$th individual ($j=1,...,J$) from a multinomial: 
$$P(\zeta_j = h|-) \propto \pi_h^* \prod_{i=1}^{n_j} \sum_{l=1}^L w^*_{lh} \prod_{m=1}^M \mathcal{K}_m(c_{ji}^{(m)}| \theta^{(m)^*}_{lh}). $$

\item Sample the membership indicator $\xi_{ji}$, for $j=1,...,J$ and $i=1,...,n_j$, with:
$$P(\xi_{ji} = l | -) \propto w_{l\zeta_j}^* \prod_{m=1}^M \mathcal{K}(c_{ji}^{(m)}|\theta_{l\zeta_j}^{(m)*} ).$$

\item Sample $\pi^*_h$ by first sampling $(u_h^*|-) \sim \text{beta}(1+m_h, \alpha + \sum_{s=h+1}^K m_s)$, $h = 1,..., K-1$, and $u_K^* = 1,$
where $m_h$ is the number of subjects assigned to cluster $h$, and then let $\pi_h^* = u_h^* \prod_{s=1}^{h-1}(1-u_s^*). $


\item Sample $w_{lh}^*$ by first sampling $(v_{lh}^*|-) \sim \text{beta}(1+n_{lh}, \beta + \sum_{s=l+1}^L n_{sh})$, $l = 1,\ldots, L-1$, $h = 1,\ldots,K$ and  
$v_{LK}^* = 1,$ where $n_{lh}$ is the number of observations assigned to atom $l$ of distribution $h$, and then $w_{lh}^* = v_{lh}^* \prod_{s=1}^{l-1}(1 - v^*_{sh}).$


\item Sample the parameters $\theta_{lh}^{(m)*}$ for $l=1,...,L$, $h=1,...,K$ and $m=1,...,M$ from 
$$P(\theta_{lh}^{(m)*}|-) \propto P_0^m(\theta_{lh}^{(m)*}) \left( \prod_{\{i,j| \zeta_j=h,\xi_{ij} = l\}} \mathcal{K}_m(c_{ji}^{(m)}|\theta_{lh}^{(m)*}) \right),  $$
where $P_0^m(\cdot)$ is the conjugate prior for parameters in $\mathcal{K}_m(\cdot|\theta^{(m)})$. If no observation is assigned to the $lh$th cluster, we draw $\theta_{lh}^{(m)*}$ from the prior $P_0^m$. 

\item Sample the concentration parameters $\alpha$ and $\beta$: 
We choose conjugate priors: $\alpha \sim \text{gamma}(a_\alpha,b_\alpha)$ and $\beta \sim \text{gamma}(a_\beta,b_\beta)$. The posterior samples for $\alpha$ and $\beta$ are constructed as
$$P(\alpha|-) \sim \text{gamma}\bigg(a_\alpha + (K-1), b_\alpha - \sum_{h=1}^{K-1} \log(1 - \mu_h^*) \bigg), $$
$$P(\beta-) \sim \text{gamma}\bigg( a_\beta + K(L-1), b_\beta - \sum_{l=1}^{L-1}\sum_{h=1}^{K} \log(1 - v_{lh}^*) \bigg). $$
\end{enumerate}
We will evaluate the performance of this Gibbs sampler through application to human brain connectome data.

 \subsection{Jointly model fiber curves and connection strength} \label{sec:jointlyfc}

Model (\ref{eqn:ndp}) does not incorporate information on the strength of connection $w_j$ between the two ROIs within individual $j$.  However, it is straightforward to generalize the model to include this additional information by letting
$$ G_j \sim \sum_{h=1}^\infty \pi_h^* \delta_{\{G_h^*(\cdot),\psi_h\}}, $$
where now the $h$th component of $G_j$ includes not only the mixing measure $G_h^*$ characterizing the distribution of fiber curves in that component but also 
parameters $\psi_h$ within a kernel $\mathcal{K}_w( \cdot; \psi_h )$ for the measure of connection strength.  The resulting joint model characterizes flexible dependence in the connection strength and fiber curves through shared dependence on the individual's cluster allocation.  For continuous measures of connection strength $w_j$, we can simply use a Gaussian kernel.  However, we will focus on $w_j$ equal to the number of connections between the regions of interest, so that $\mathcal{K}_w( w; \psi)$ is a parametric distribution with support on the non-negative integers.  To induce this kernel, we apply the approach of \citet{Canale2011} and simply `round' a Gaussian kernel with unknown mean and variance, with negative values mapped to 0, values in (0,1) mapped to 1, values in (1,2) mapped to 2 and so on.  Posterior sampling can proceed via a slight modification of the sampler of Section \ref{sec:pp_ps}, with details provided in a Supplement.

\section{Application to human brain connectome data}
We consider two data sets: a Test-Retest Dataset and the Human Connectome Project Dataset. 

{\bf Test-Retest Dataset}:  Contains 3 scans for each subject taken at one month intervals.  A total of 15 acquisitions, from 5 healthy participants, were utilized for our analysis. In each scan, a dMRI image and an anatomical T1-weighted image were acquired on a 1.5 Tesla SIEMENS Magnetom.  The dMRI image has a 2$mm$ isotropic resolution and was acquired along $64$ uniformly distributed directions. The T1 image has a 1$mm$ isotropic resolution. Diffusion data was up-sampled (using a trilinear interpolation) to the same resolution as T1 image before performing tractography. The T1 image was parcellated using Freesurfer (with Desikan-Killiany atlas) and registered to the diffusion domain. Quality control by manual inspection was used to verify the parcellation and registration. 

{\bf Human Connectome Project (HCP) Dataset}: The 2016 HCP data contain about $900$ subjects, and we focus on the $857$ subjects having both diffusion data and an anatomical T1-weighted image.  The dMRI images in HCP have isotropic voxel size of $1.25mm$, and $270$ diffusion weighted scans. HCP has processed the diffusion image and T1 image such that they have the same resolution and lie in the same space (aligned).  Desikan-Killiany parcellation for each T1 image is also provided. See \cite{VanEssen20122222} for more details about the HPC data.  

The tractography dataset for each subject is generated using the probabilistic method of \citet{Girard2014266} with the recommended optimal parameters. Each streamline in the constructed tractography has a step size of 0.2$mm$. About 1 million fiber curves for each subject are generated. Under the Desikan-Killiany atlas, the brain cortical bands are segmented into $68$ anatomical regions (34 regions per hemisphere). The fiber curves connecting any pair of regions are extracted. Before applying our method to each connection, outlying fiber curves that do not follow the major white matter pathways (false positives caused by the fiber tracking algorithm) are removed using a similar method proposed by \cite{Cote2015}.  

\subsection{Component estimation}
\begin{figure}
\begin{center}
\begin{tabular}{c}
\includegraphics[height=2.6in]{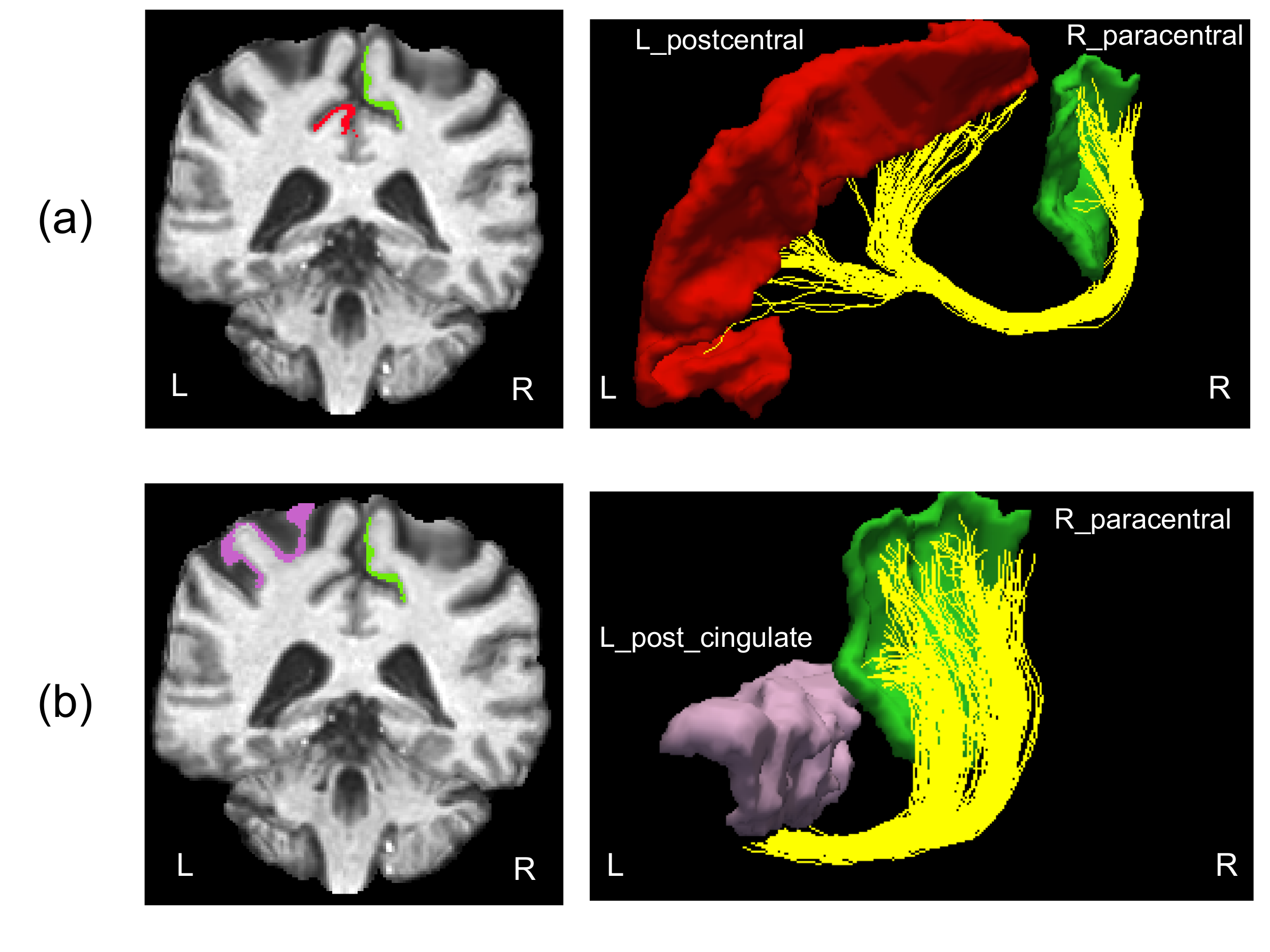}
\end{tabular}
\caption{Example of two connections we used in this paper: (a) between {\it r\_pl} and {\it l\_pg} and (b) between  {\it r\_pl} and {\it l\_pcc}. } \label{fig:twocon}
\end{center}
\end{figure}

For any two regions $(r_a, r_b)$, to learn a low-dimensional structure $L_{(r_a,r_b)}$ representing the shape component, we randomly selected $30$ subjects from HCP  as the training data and learnt a set of basis functions using FPCA. We focus on two connections: (1) between {\it right paracentral lobule} ({\it r\_pl}) and {\it left postcentral gyrus} ({\it l\_pg}); (2) between  {\it right paracentral lobule} ({\it r\_pl}) and {\it left  posterior cingulate cortex} ({\it l\_pcc}). Figure \ref{fig:twocon} illustrates these connections in a subject of the test-retest dataset. 

\begin{figure}
\begin{center}
\begin{tabular}{cccc}
\includegraphics[height=1.1in]{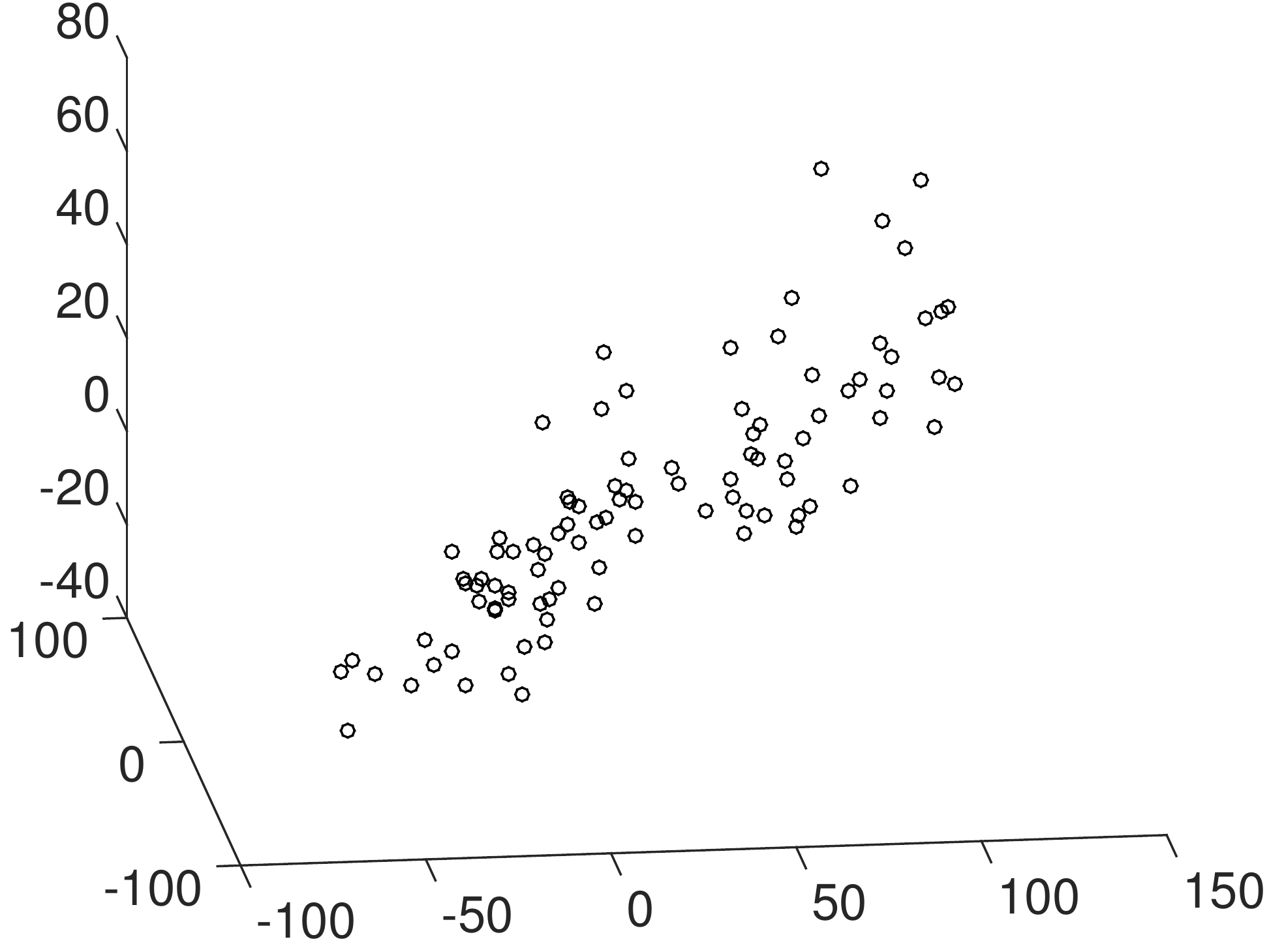} &
\includegraphics[height=1.1in]{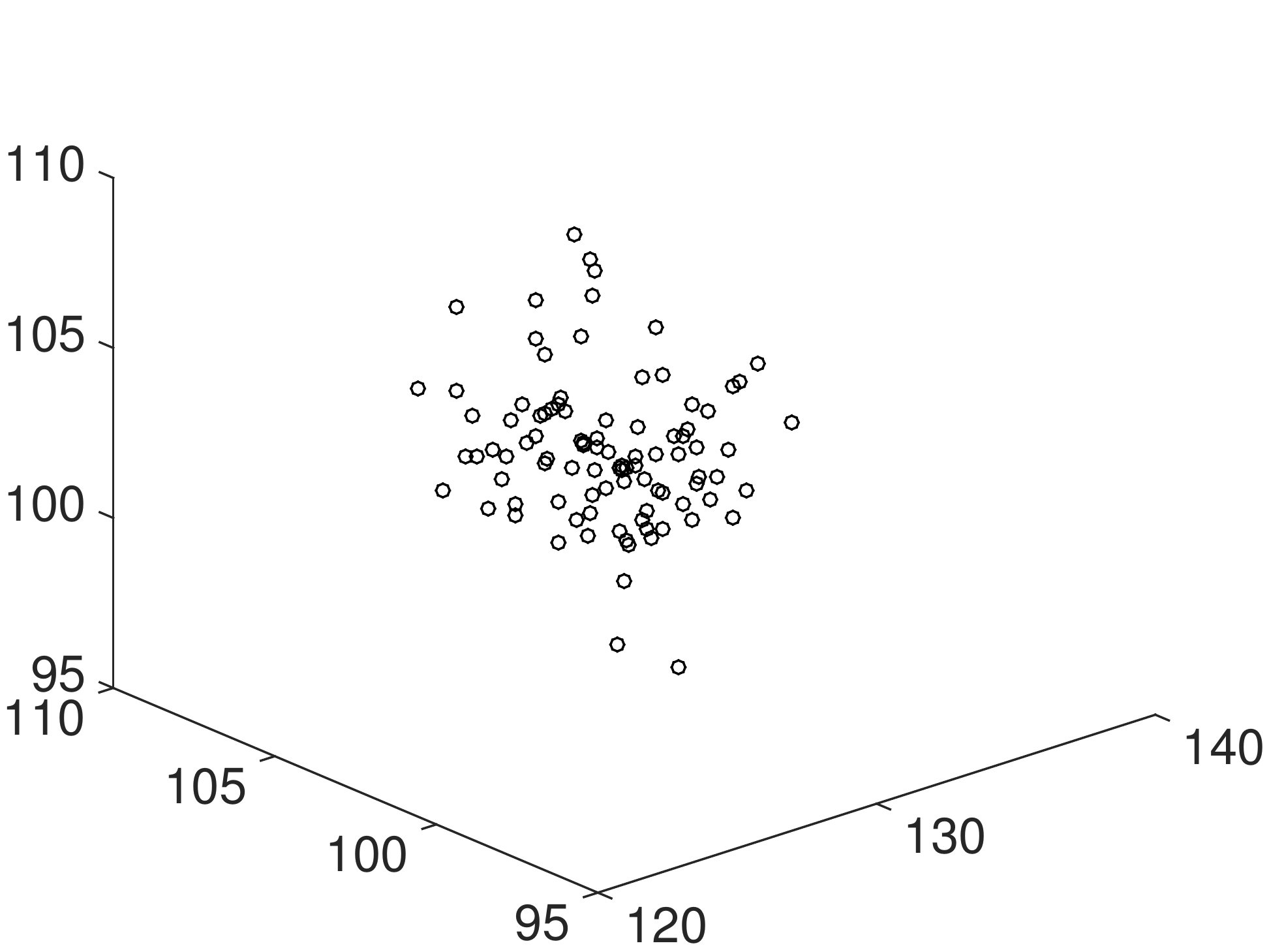} &
\includegraphics[height=1.1in]{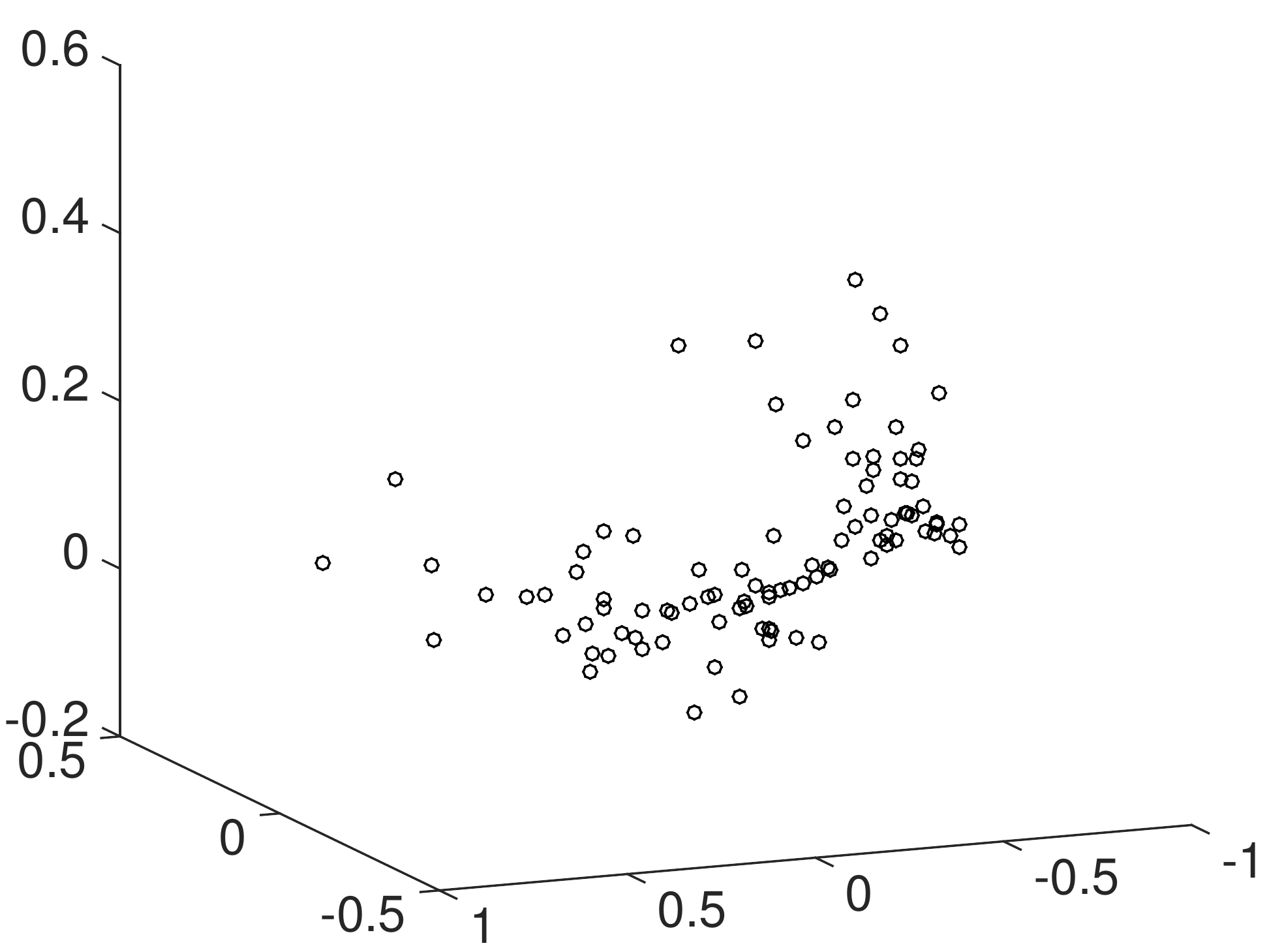} &
\includegraphics[height=1.2in]{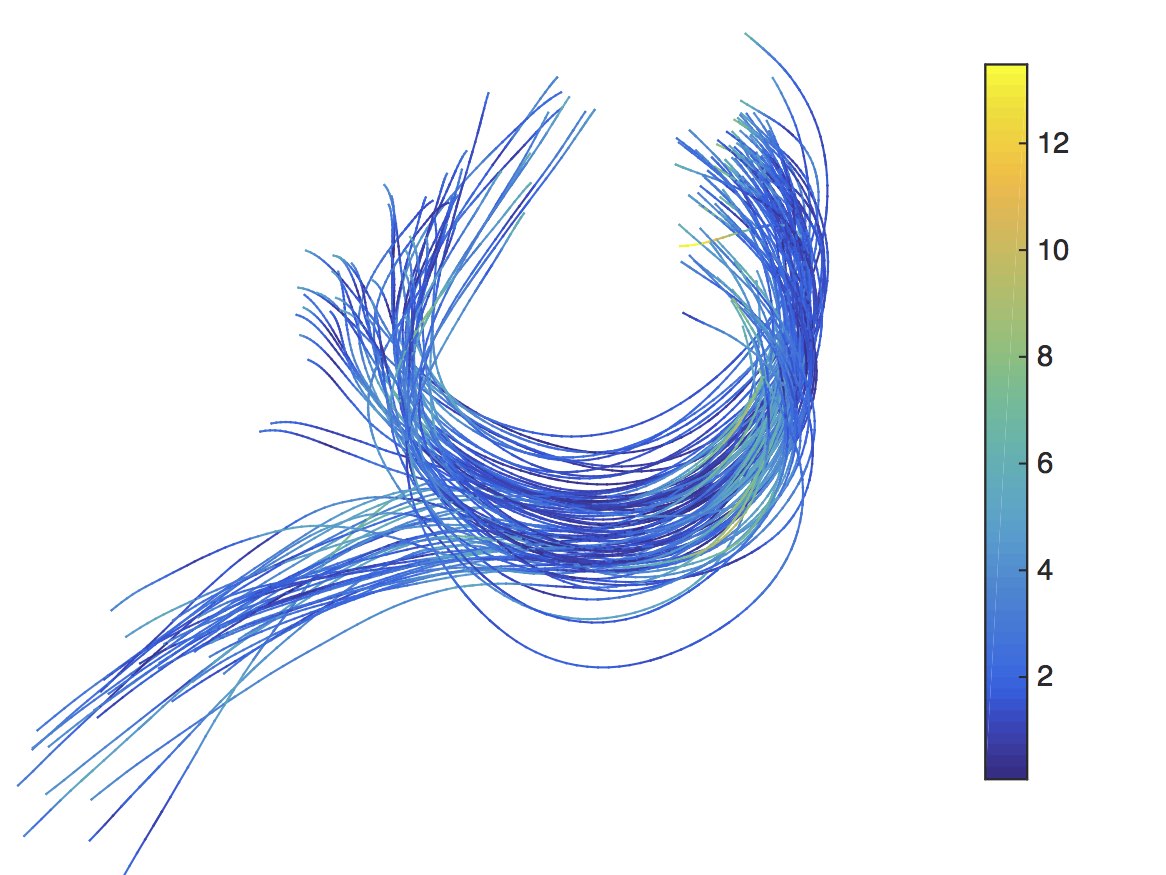} \\
\includegraphics[height=1.1in]{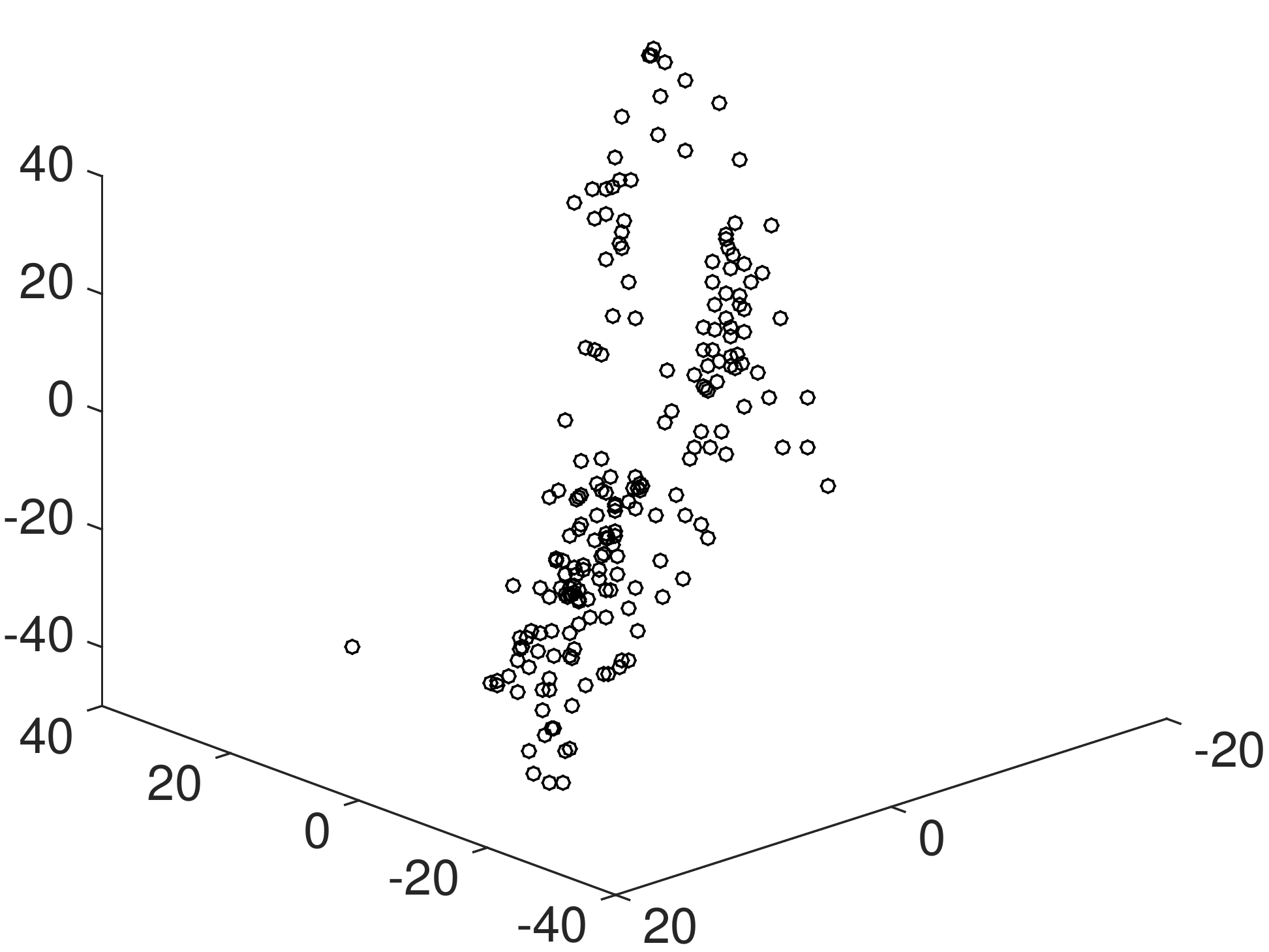} &
\includegraphics[height=1.1in]{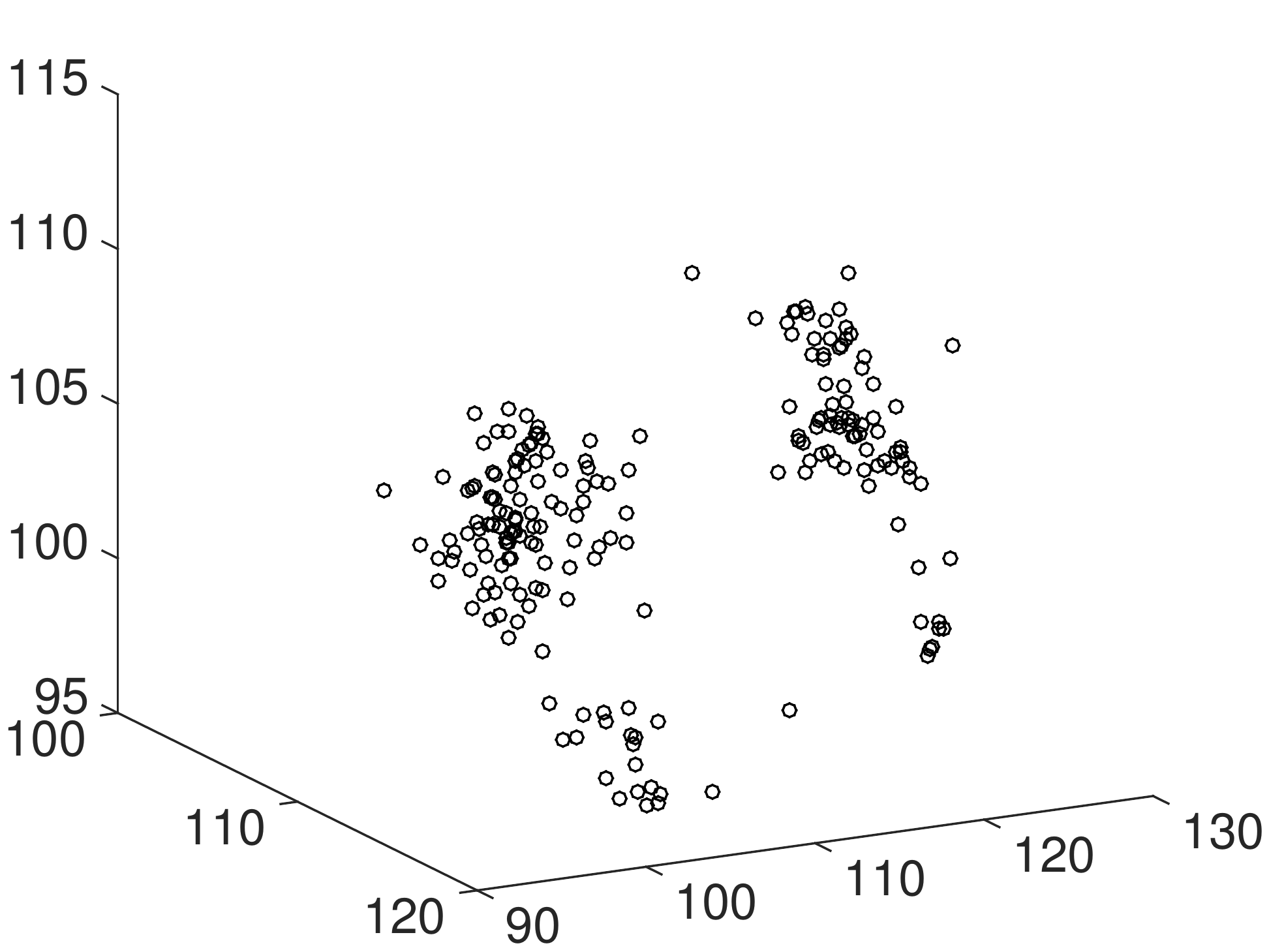} &
\includegraphics[height=1.1in]{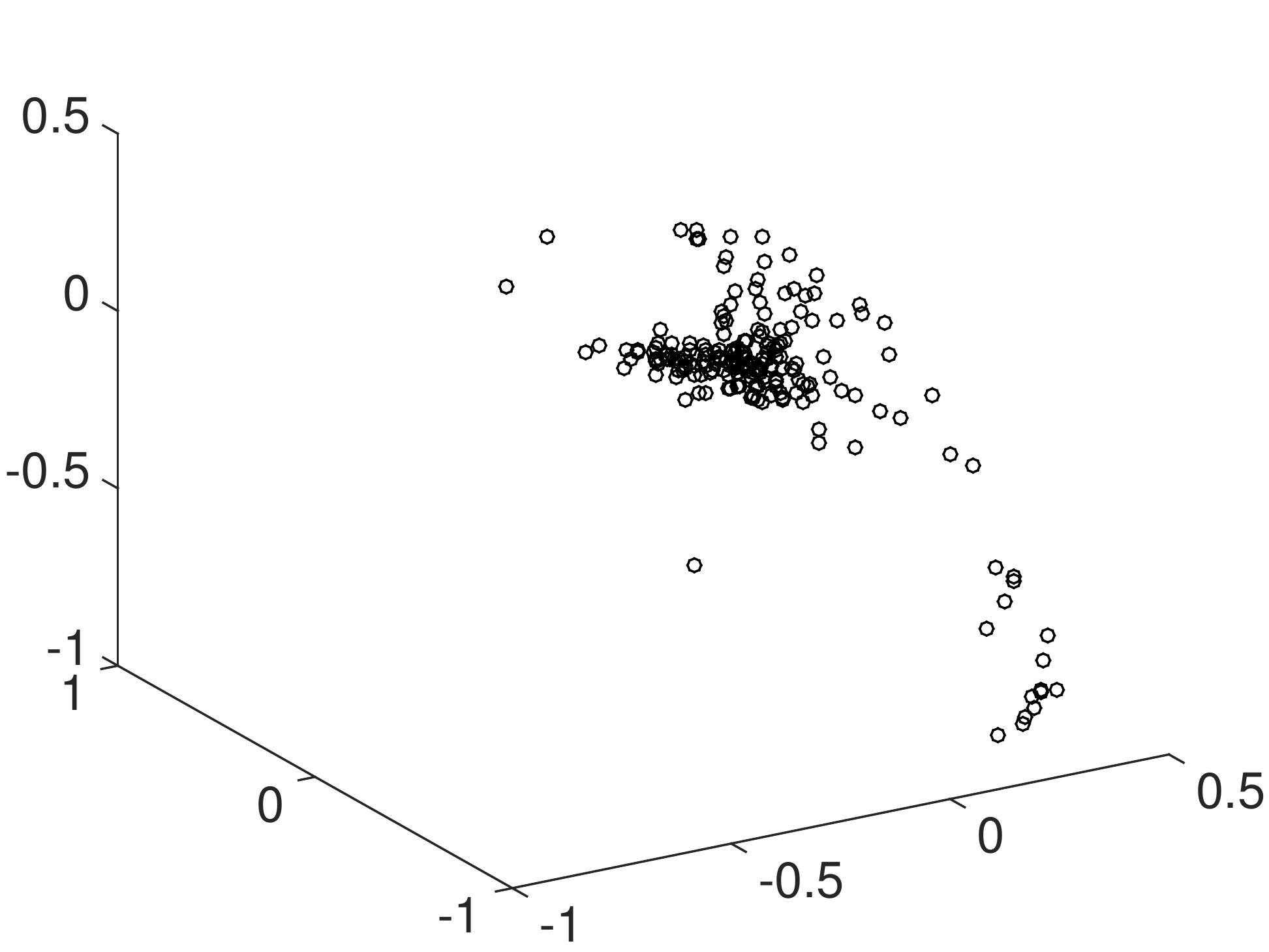} &
\includegraphics[height=1.2in]{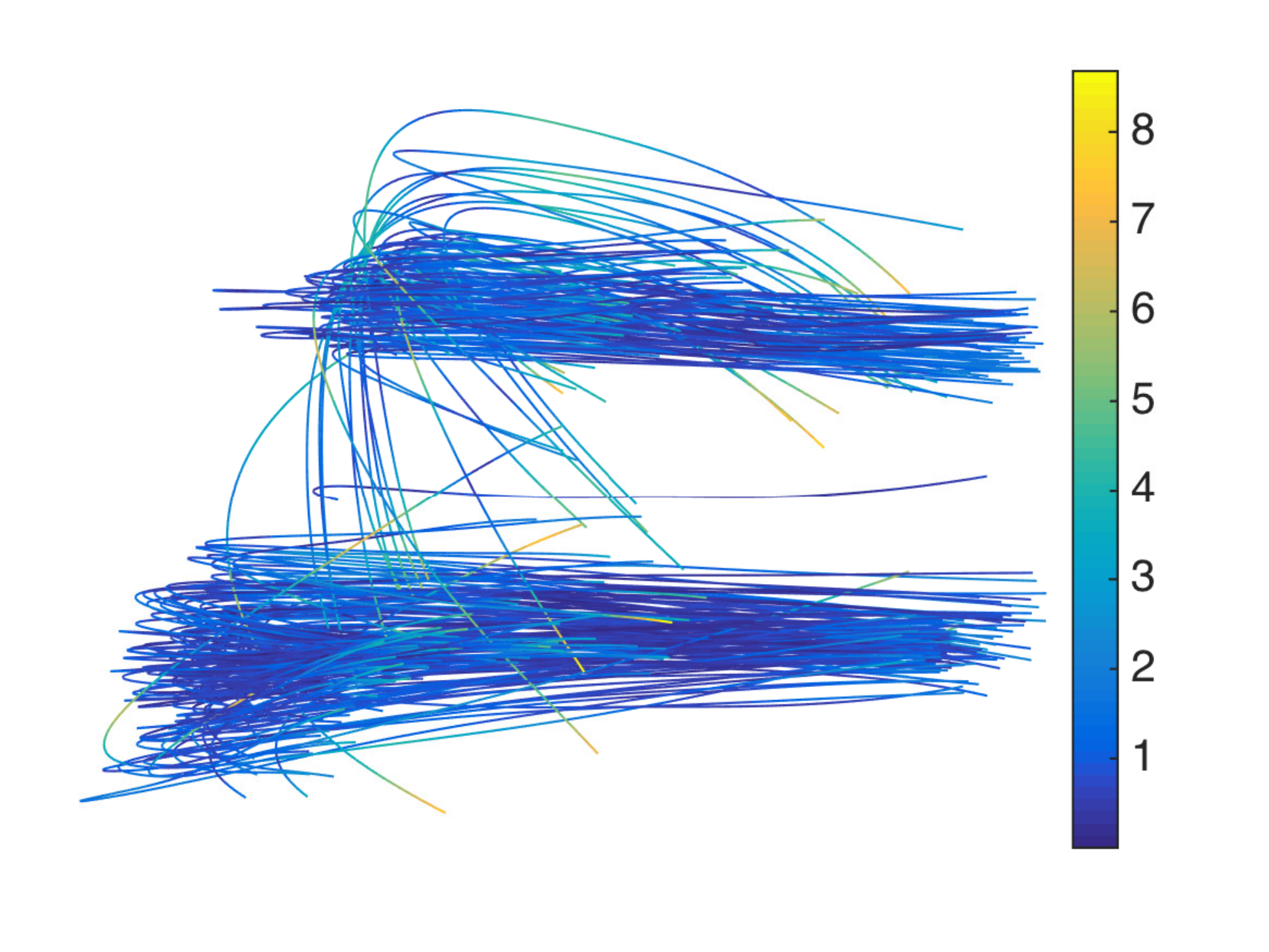} \\
(a) & (b) & (c) & (d) \\
\end{tabular}
\caption{Illustration of decomposed components and the recovered fiber curves. (a), (b) and (c) show the shape, translation and rotation components, respectively. (d) shows the recovered fiber curves using these components. The color indicates the difference (in a unit of $mm$) between the recovered and original fiber curve.  } \label{fig:comp_all}
\end{center}
\end{figure}

Using the method introduced in Section 2.3, we estimated three components $c^{(1)}$, $c^{(2)}$ and $c^{(3)}$ for all fibers. For $c^{(2)} \in \Re^T$, we set $T=3$, so we use three coefficients (on three major FPCA basis functions) to represent a fiber. Using a larger $T$ will increase representation precision, but we found $T=3$  is sufficient. In Figure \ref{fig:comp_all}, we plot the estimated components for fiber curves in the two connections (shown in Figure \ref{fig:twocon}). For the connection ({\it r\_pl}, {\it l\_pg}), the fiber curves start from the right paracentral lobule, group into a bundle, traverse the  corpus callosum, and then split into two bundles to connect the left postcentral gyrus region. The split makes the fiber curve have two distinct shapes, and therefore, we expect that the shape component should be able to tell this split.  For the connection ({\it r\_pl}, {\it l\_pcc}),  there are a few distinct pathways, differing in both shape and location. Therefore, the shape and translation components should contain the most geometric information about this connection. The rotation components in both cases center around the origin and it is unclear how much information they have.  In Figure \ref{fig:comp_all} (d), we plot the recovered fiber curves using  $c^{(1)}$, $c^{(2)}$ and $c^{(3)}$.  The color along the curves indicates the discrepancy (with a unit of $mm$) between the original fiber and the recovered fiber. We can see that with only $9$ parameters, we can accurately recover most fibers. The biggest discrepancy generally focuses on the starting and ending points. The main reason is that the starting and ending points are either in the gray matter or in the interface of gray matter and white matter. Diffusion is close to isotropy \citep{Descoteaux2009} in these regions, which makes accurate fiber reconstruction intrinsically difficult. 

\subsection{Model for individual's brain connection}
We applied the model in Section 3 for the fiber curves in each connection. We are interested in answering two questions: (a) among the components (shape, translation and rotation), which one contains the most geometric information about a connection; (b) can the proposed model efficiently capture the geometric information? 

We first used the nonparametric mixture model defined in (\ref{eqn:ind}) to explore the geometric information inside each component separately.   The multivariate data  were centered to the origin and rescaled such that each coordinate has unit variation. The prior specification and posterior sampling procedure are described in Section 3.3. We assigned a normal-inverse-Wishart NIW$(\mu_0^{(m)},\lambda_0^{(m)},\Phi_0^{(m)},v_0^{(m)})$ for $P_0^m(\theta^{(m)})$ $(\theta^{(m)} = \{\mu^{(m)},{\bf \Sigma}^{(m)}\}$), where $\mu_0^{(m)} = [0,0,0]^T, \lambda_0^{(m)}=1, \Phi_0^{(m)} ={\bf I}_3$, and $v_0^{(m)} = 5$ for $m=1,2,3$, implying that $E(\mu^{(m)}|{\bf \Sigma}^{(m)}) = [0,0,0]^T$ and $E( {\bf \Sigma}^{(m)}) = {\bf I}_3$.  The inference is based on $10,000$ samples from the MCMC sampler after a burn in of $1,000$ samples. It takes about 4 minutes to draw $10,000$ samples using our MATLAB implementation with a 2.5 GHz Intel Core i7 CPU.  The results are robust to small to moderate changes in prior specification, and there is no evidence of lack of convergence. 

\begin{figure}
\begin{center}
\begin{tabular}{cccc}
\includegraphics[height=1in]{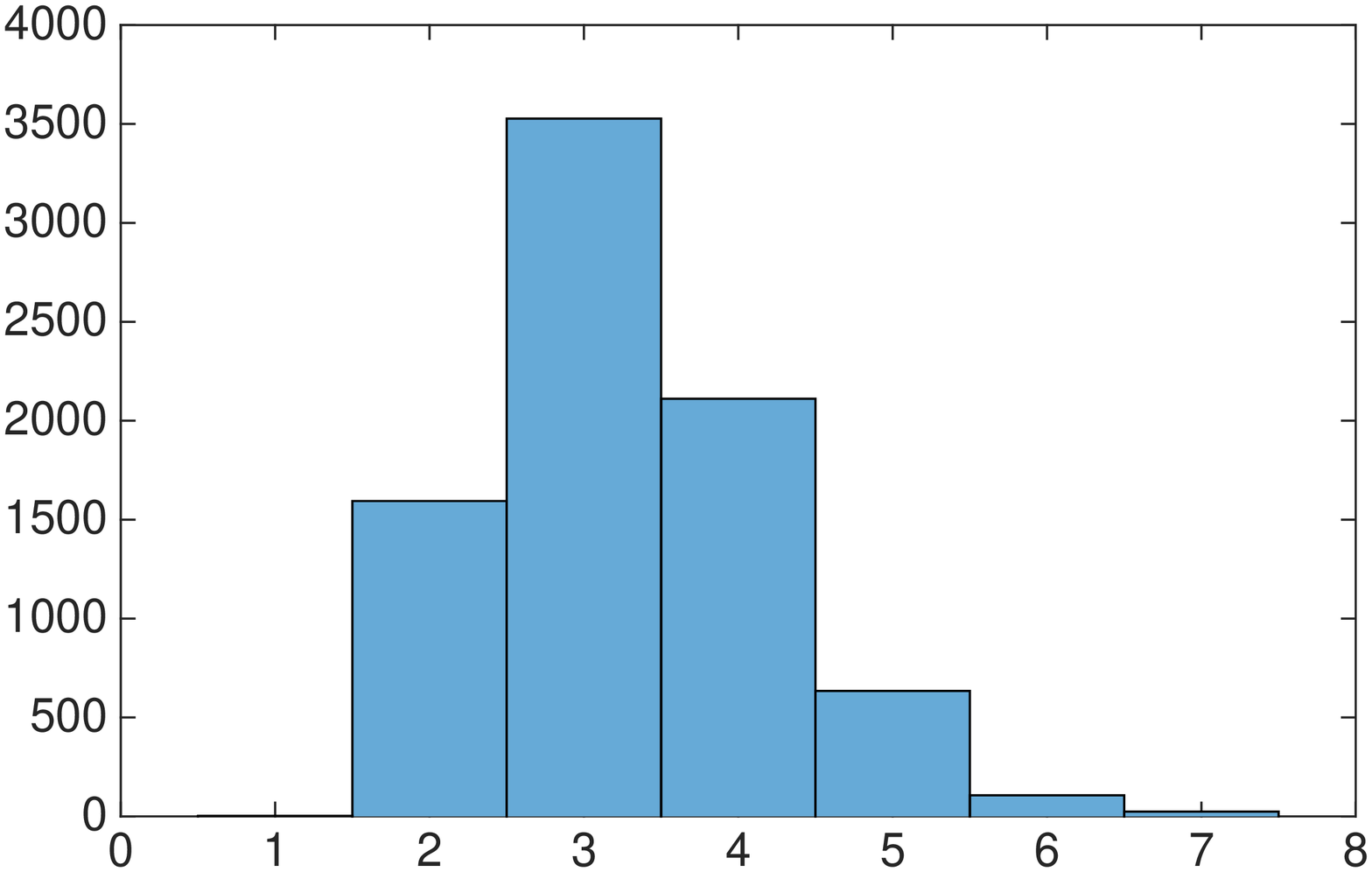} &
\includegraphics[height=1in]{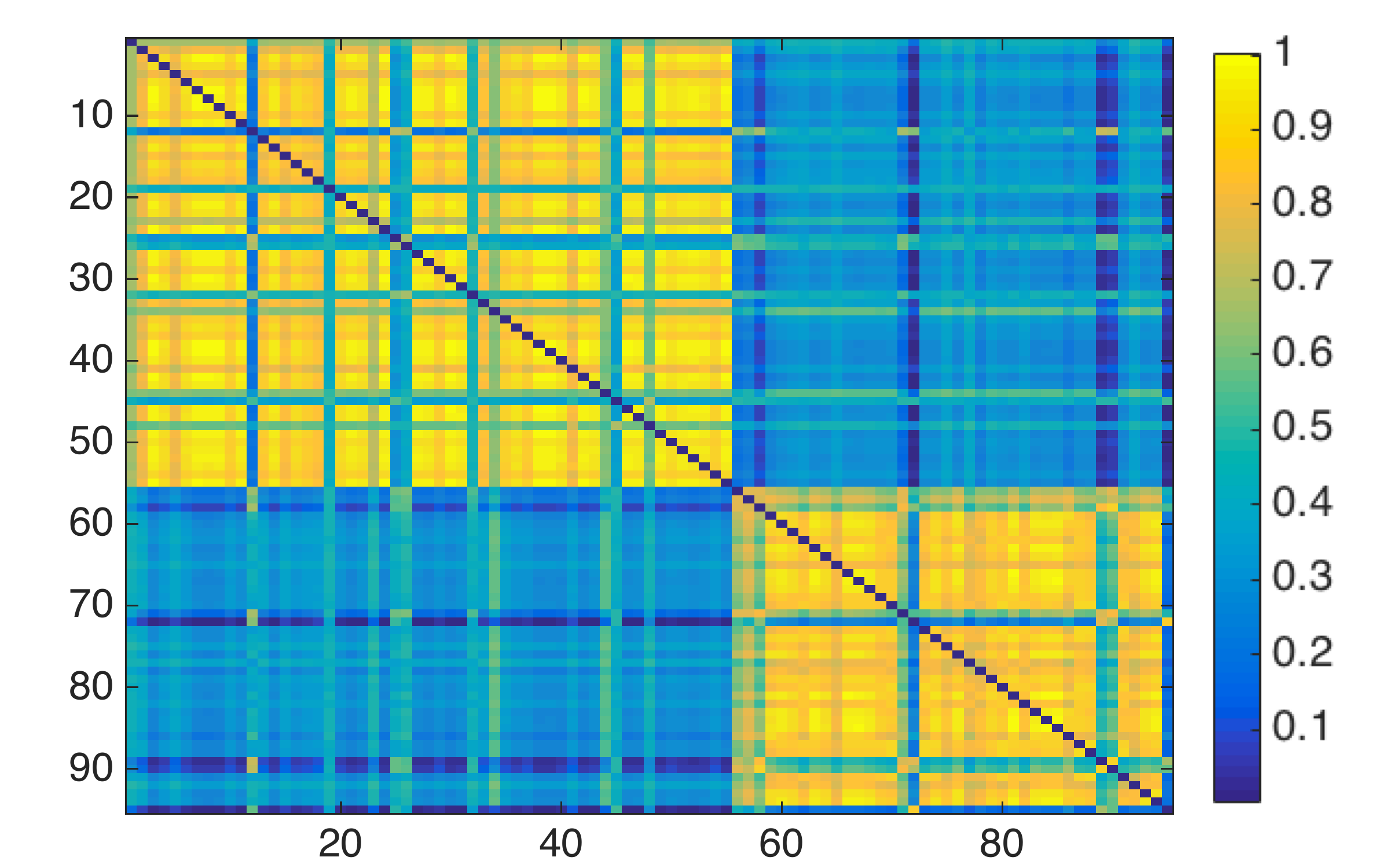} &
\includegraphics[height=1.0in]{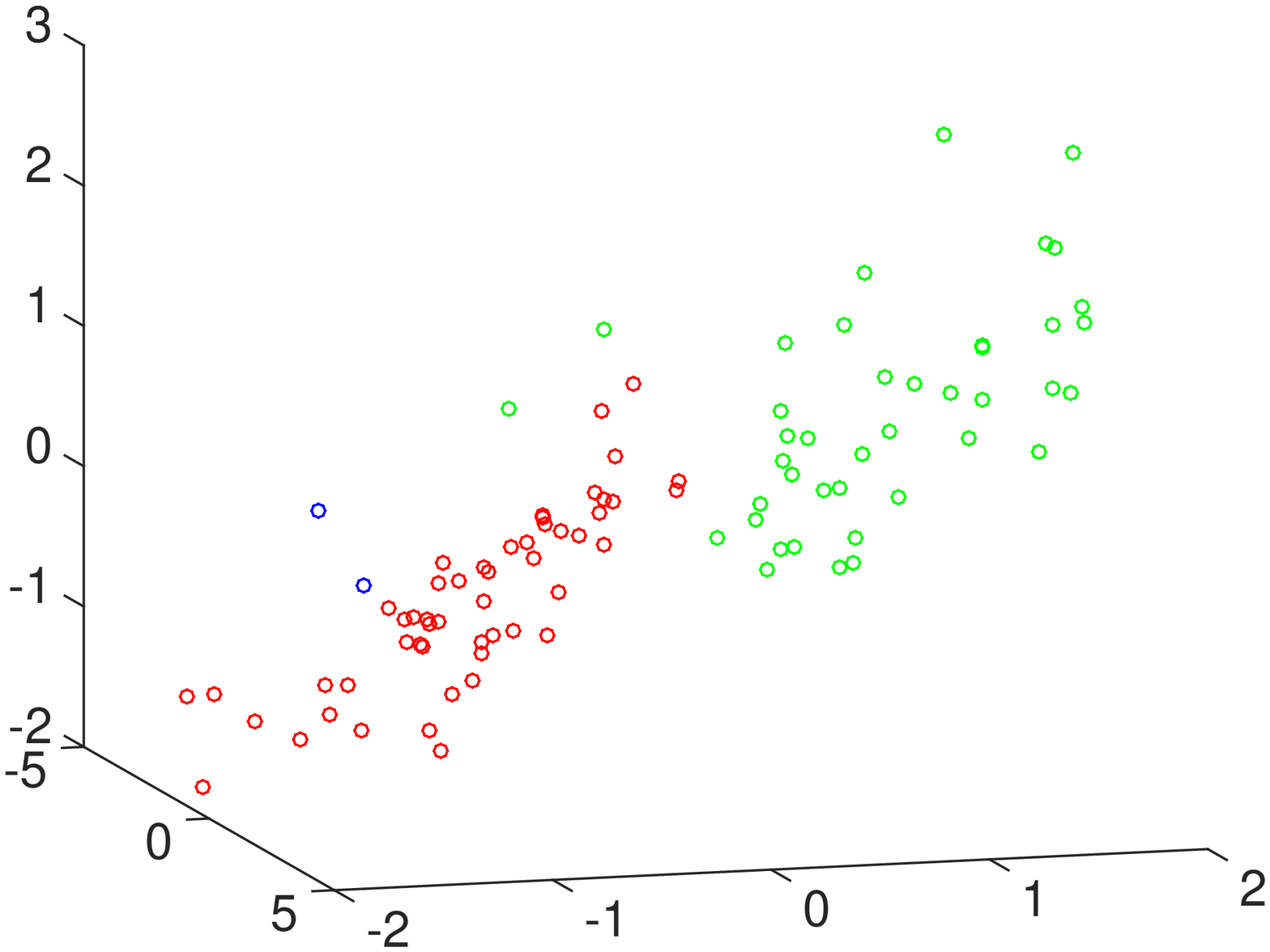} &
\includegraphics[height=1.0in]{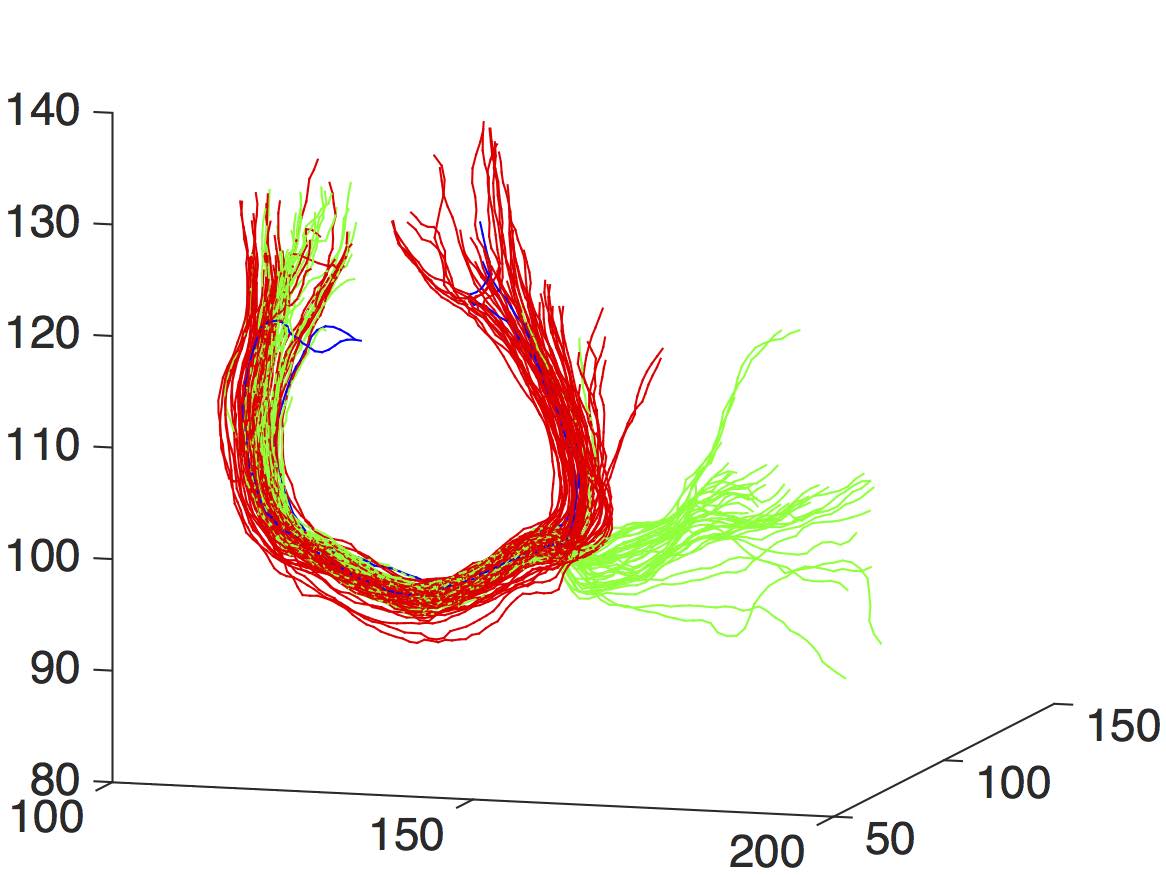} \\
\includegraphics[height=1in]{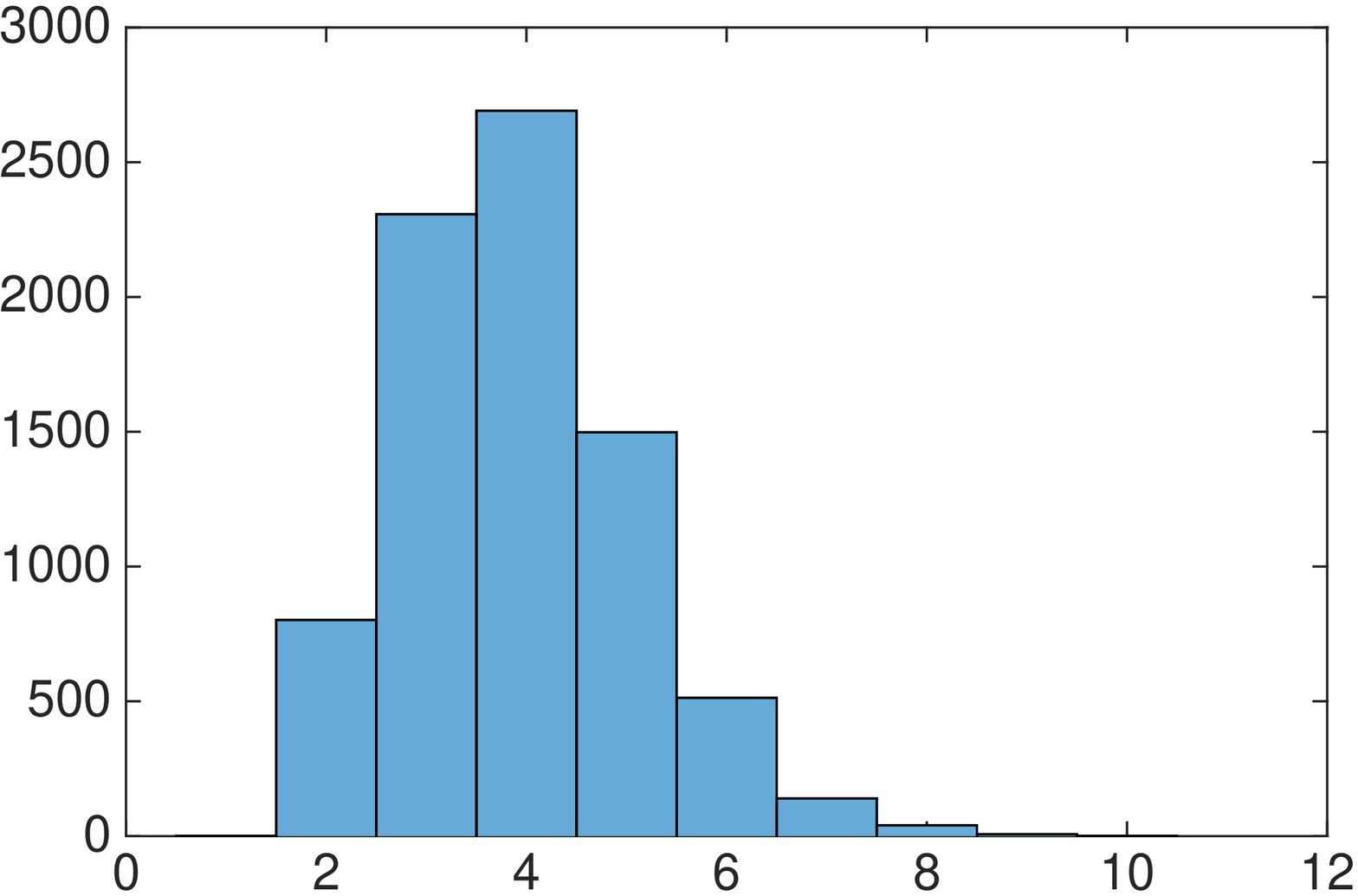} &
\includegraphics[height=1in]{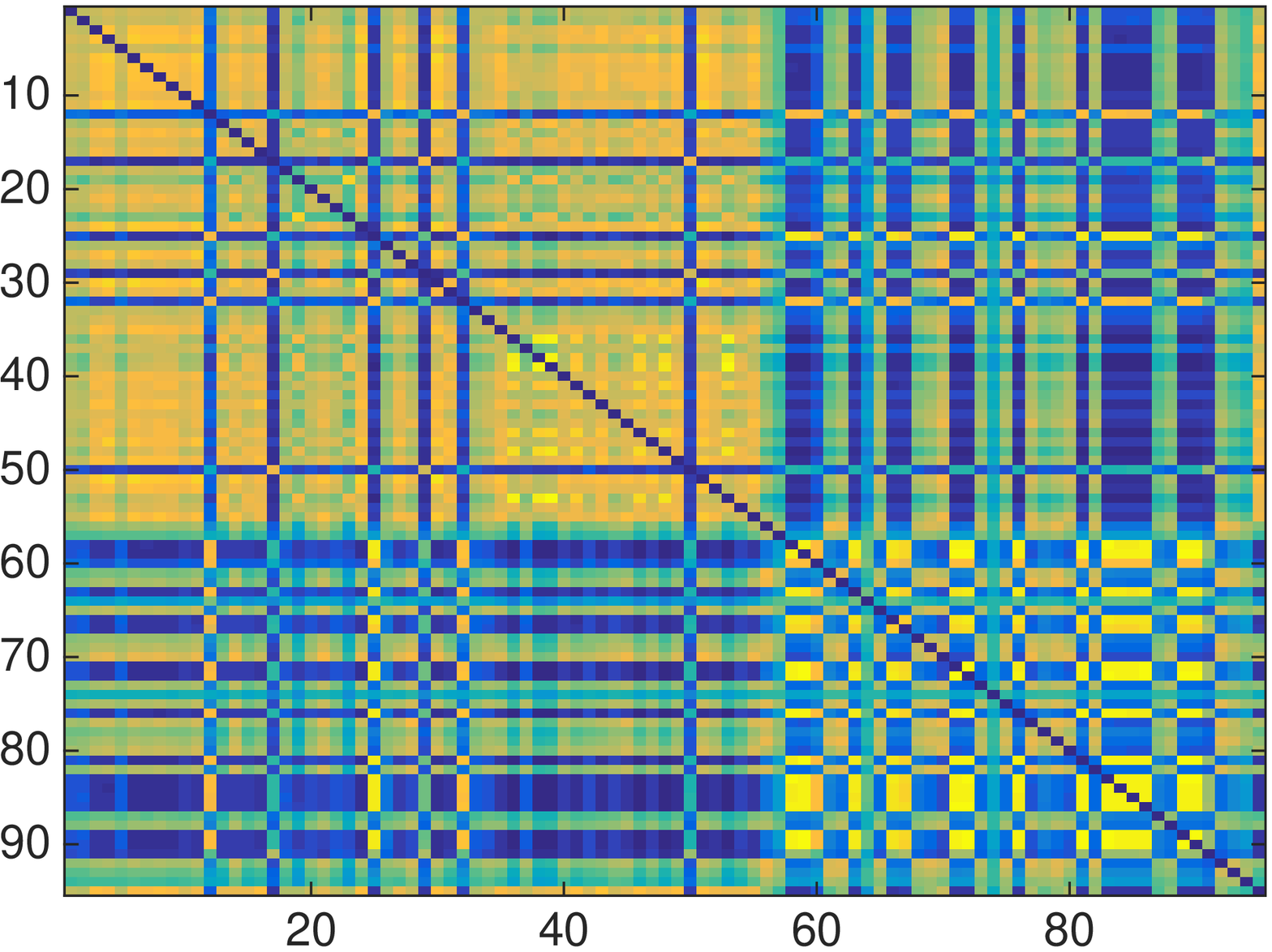} &
\includegraphics[height=1.0in]{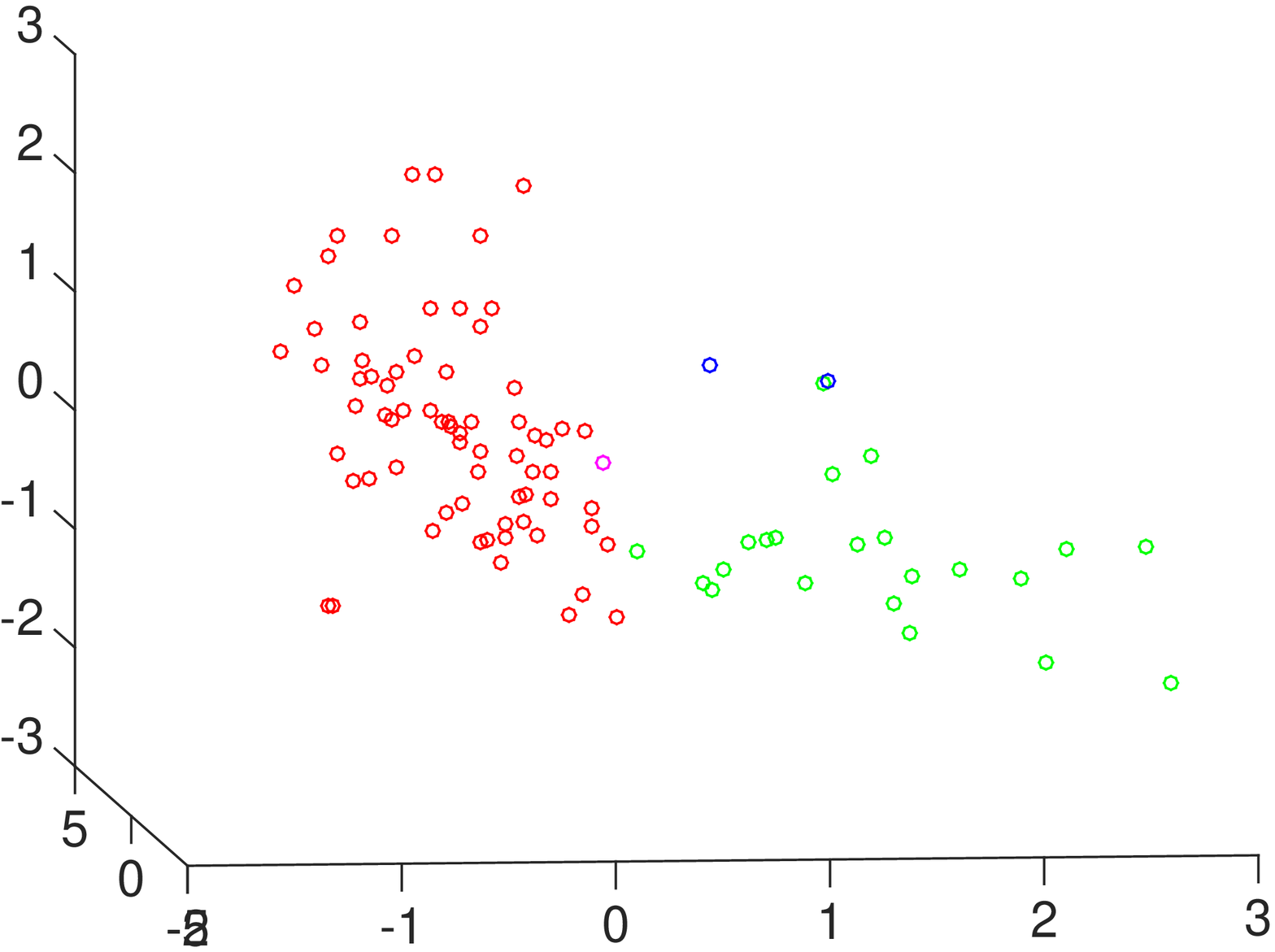} &
\includegraphics[height=1.0in]{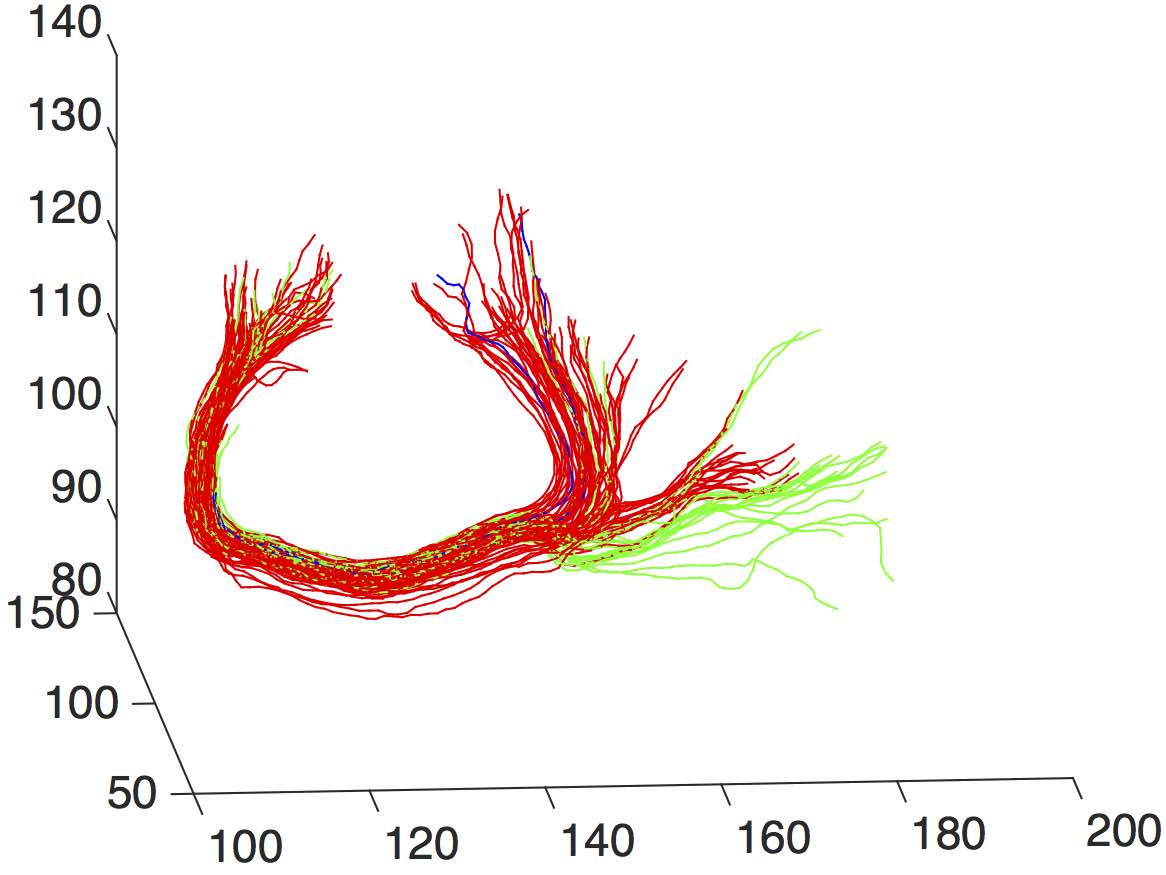} \\
\includegraphics[height=1in]{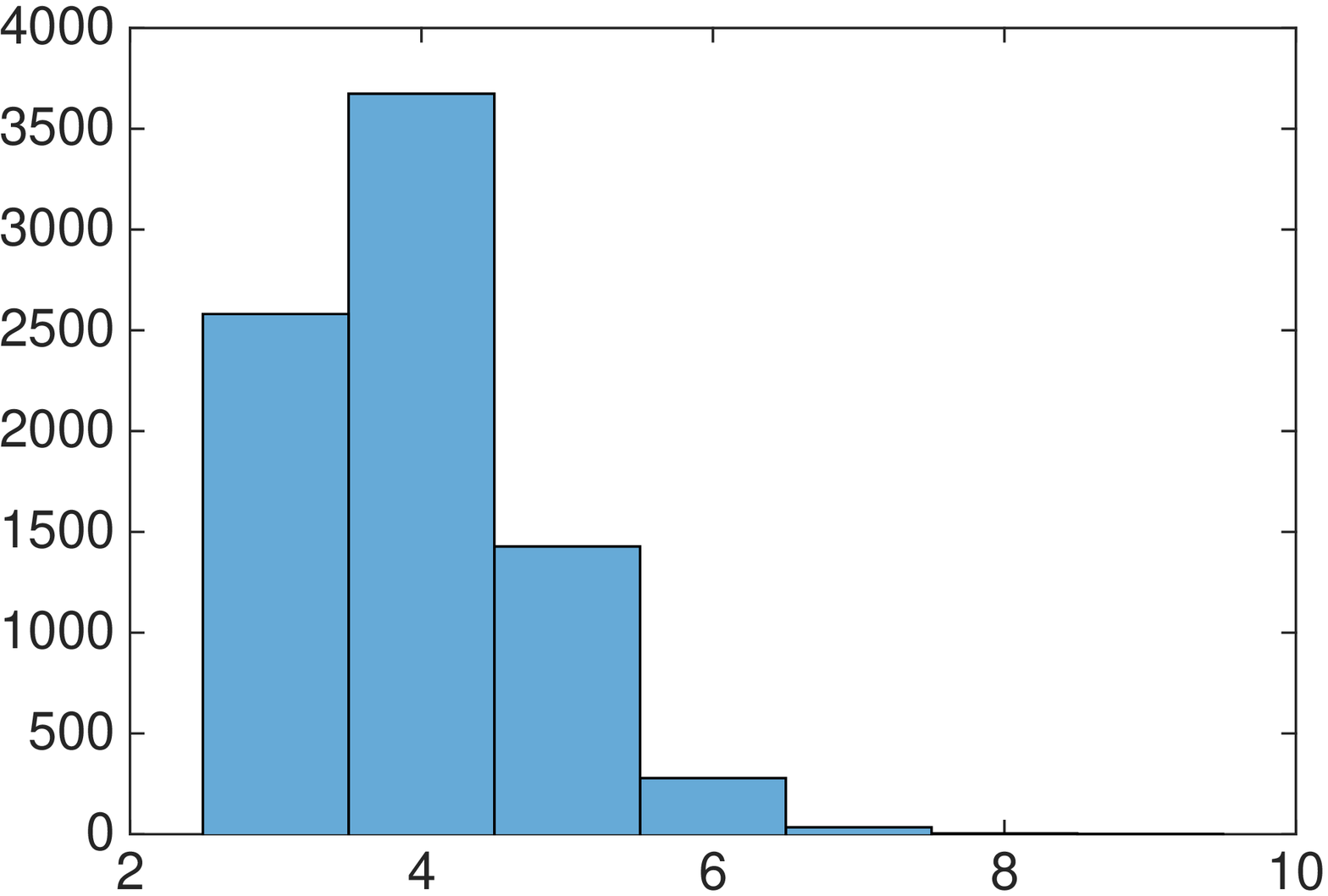} &
\includegraphics[height=1in]{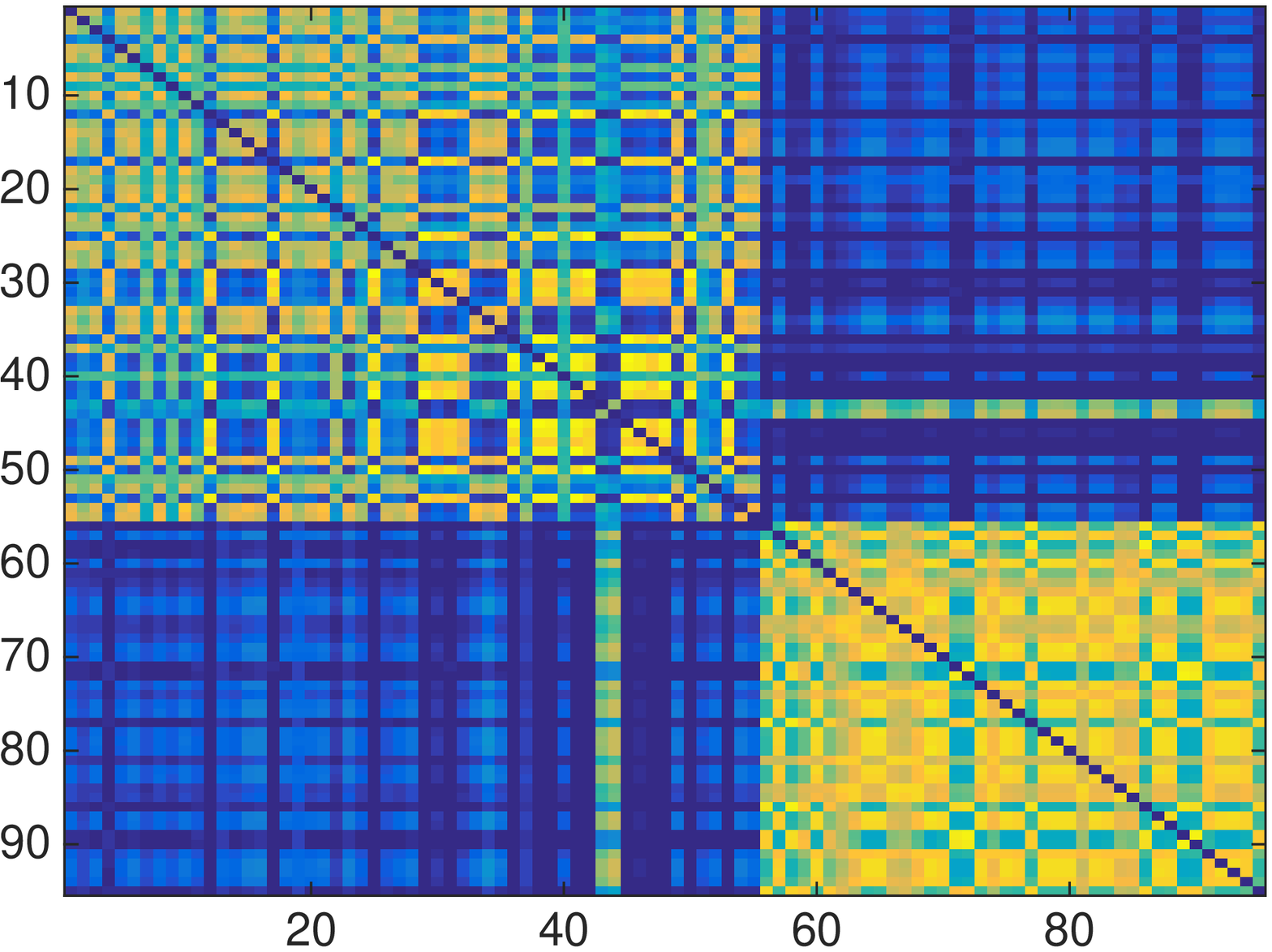} &
\includegraphics[height=1.0in]{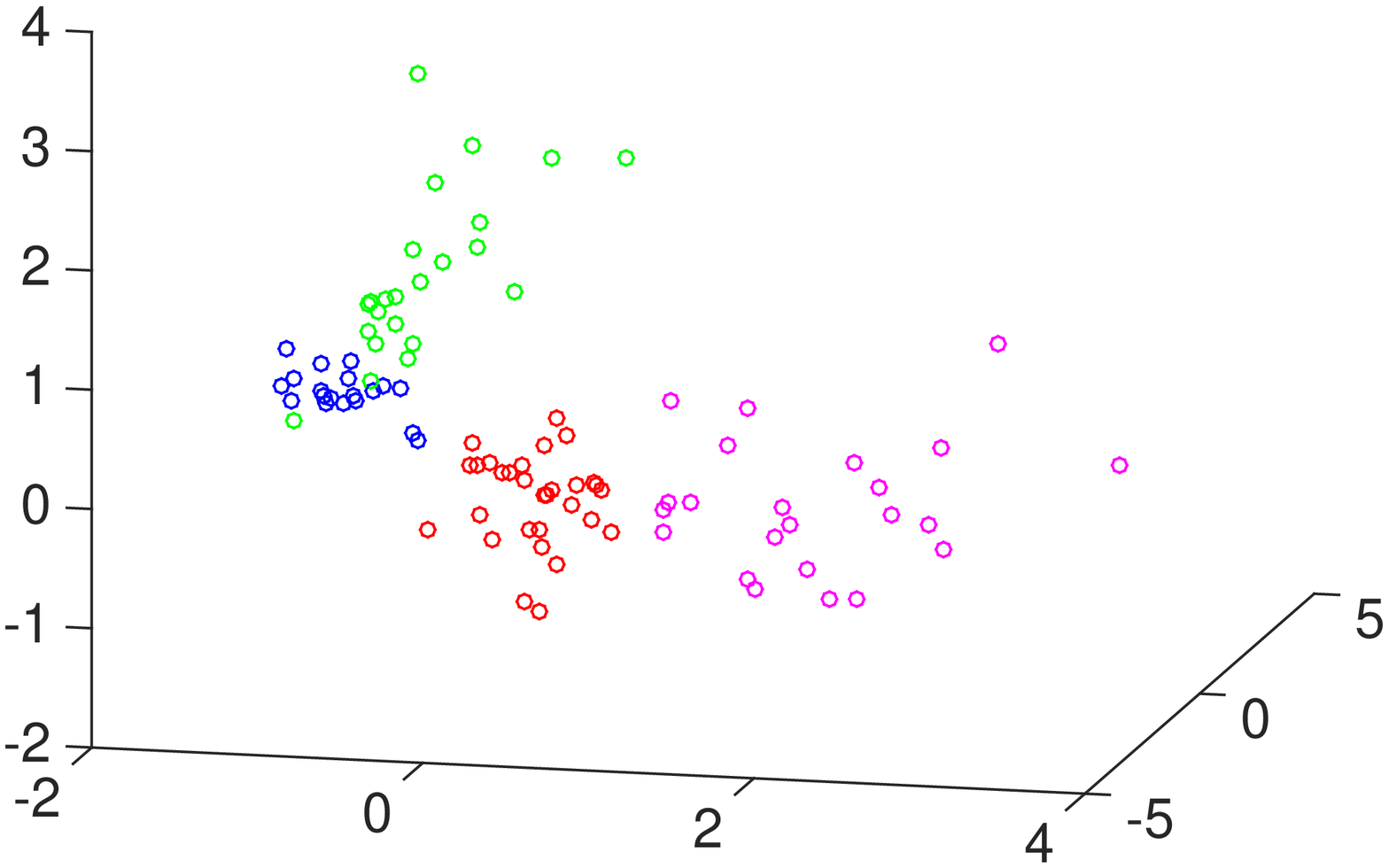} &
\includegraphics[height=1.0in]{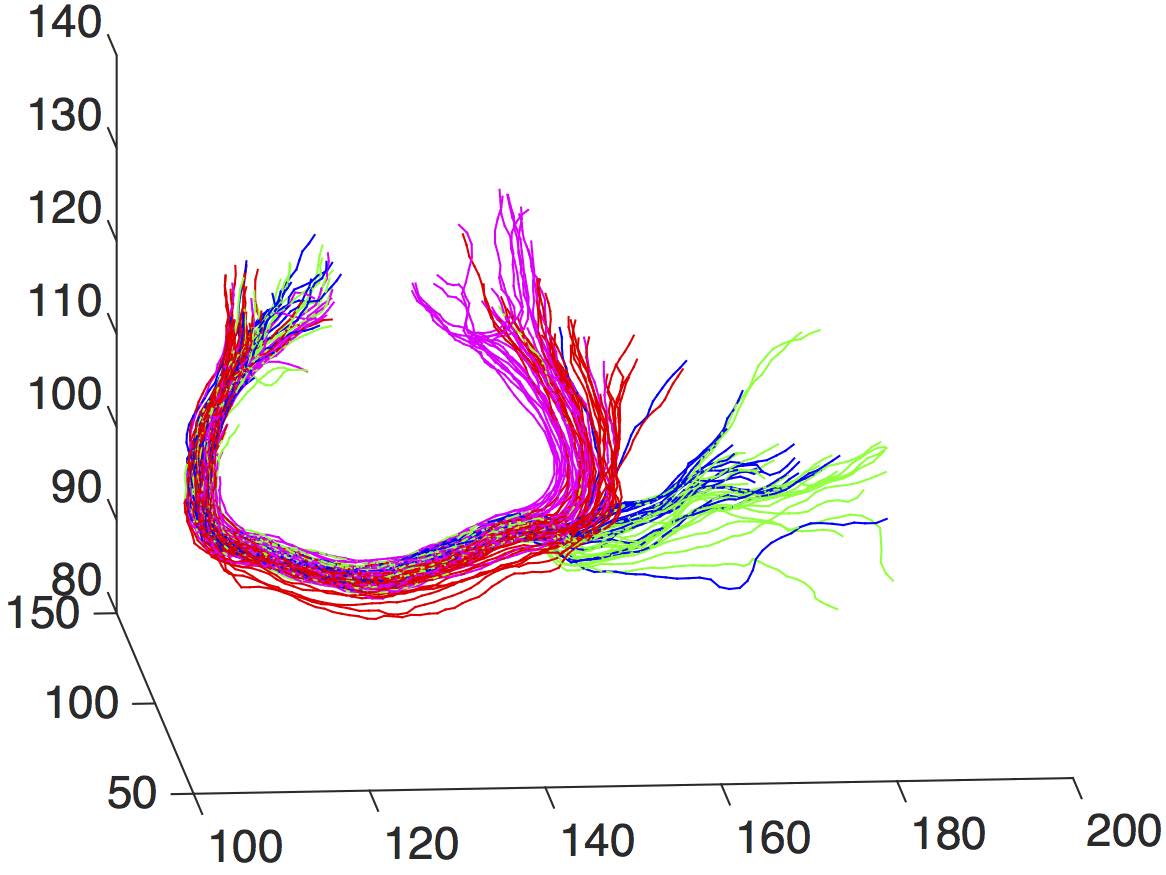} \\
(a) Posterior dist. of $k$ & (b) Adjacency matrix & (c) Clustering of $\{c_i^{(m)}\}$ & (d) Clustering of fibers  \\
\end{tabular}
\caption{Posterior summary for the connection ({\it r\_pl}, {\it l\_pg}).   The three rows are results for the shape, translation and rotation component, respectively. } \label{fig:fiber_conn1_onecomp}
\end{center}
\end{figure}

Each component contains different geometric information about the connection and such differences are reflected in the clustering result inferred based on the posterior samples. Figure \ref{fig:fiber_conn1_onecomp} summarizes the clustering results for connection ({\it r\_pl}, {\it l\_pg}). The first row shows the result for the shape component, the second row shows the result for the translation component and the last row shows the result for the rotation component. Column (a) shows posterior samples of number of clusters and (b) shows the pairwise probability heat map according to the posterior samples. To make sense of the heat map, we reordered the fiber curves such that fibers with similar shapes are close to each other. 

There are several approaches to obtain a final clustering configuration from the pairwise probability matrix \citep{Abel2008,Zhang2015171}.  Following \cite{Zhang2015171}, we use the mode of the posterior distribution on number of clusters as the final cluster number $k$.  The final cluster configuration is estimated by mapping the pairwise probability matrix into a membership matrix minimizing the discrepancy to the pairwise probability matrix and having $k$ clusters. Figure \ref{fig:fiber_conn1_onecomp} (c) and (d) show the clustering results on $\{c_i^{(m)} \}$ and the original fibers using this method.  It is clear that shape plays the 
most important role. With only the shape component, we can distinguish the two fiber bundles inside the connection.  The translation and rotation components contain some geometric information (based on the final clustering results), but much less than shape. 

\begin{figure}
\begin{center}
\begin{tabular}{ccc}
\includegraphics[height=1.2in]{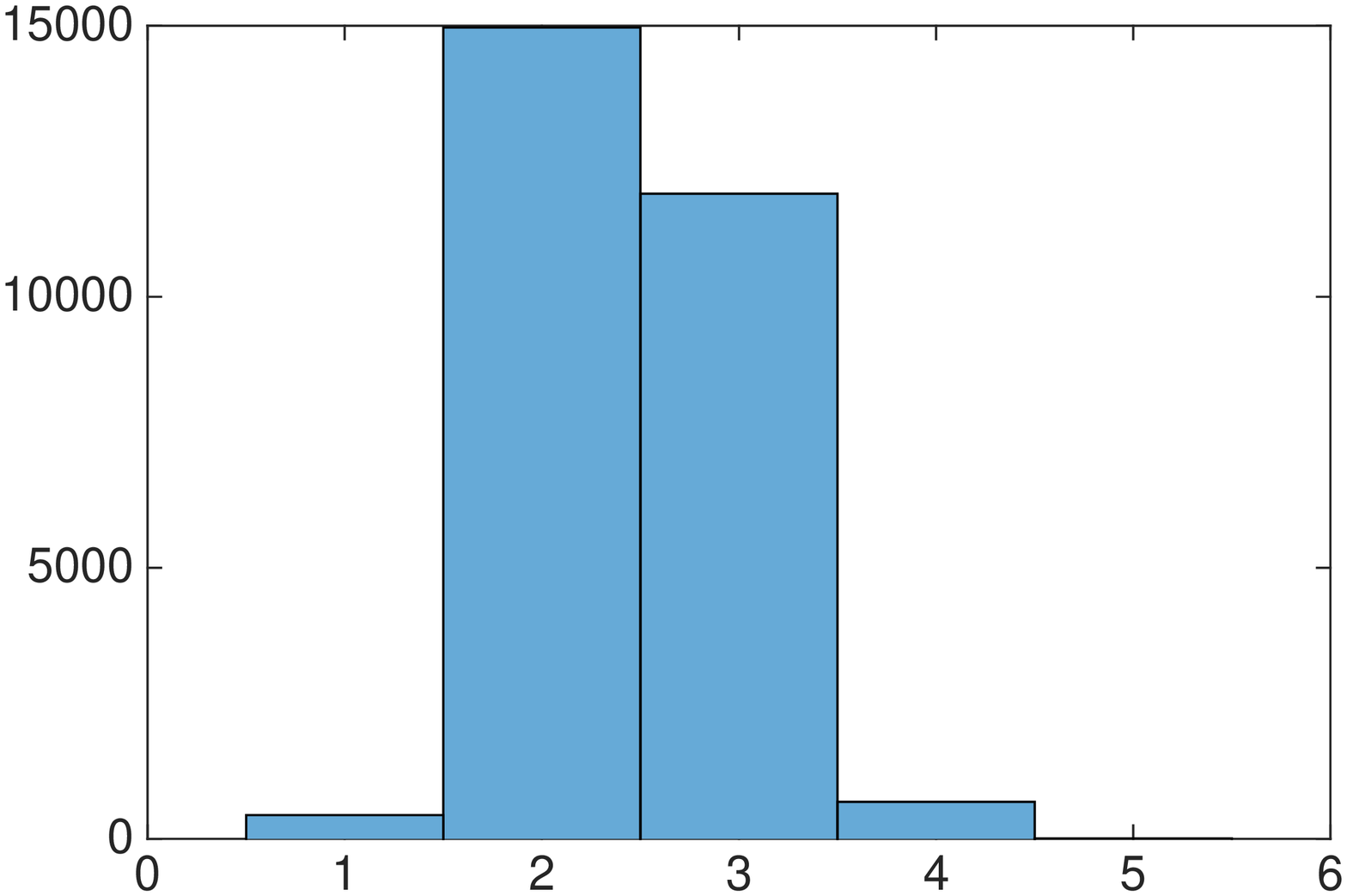} &
\includegraphics[height=1.2in]{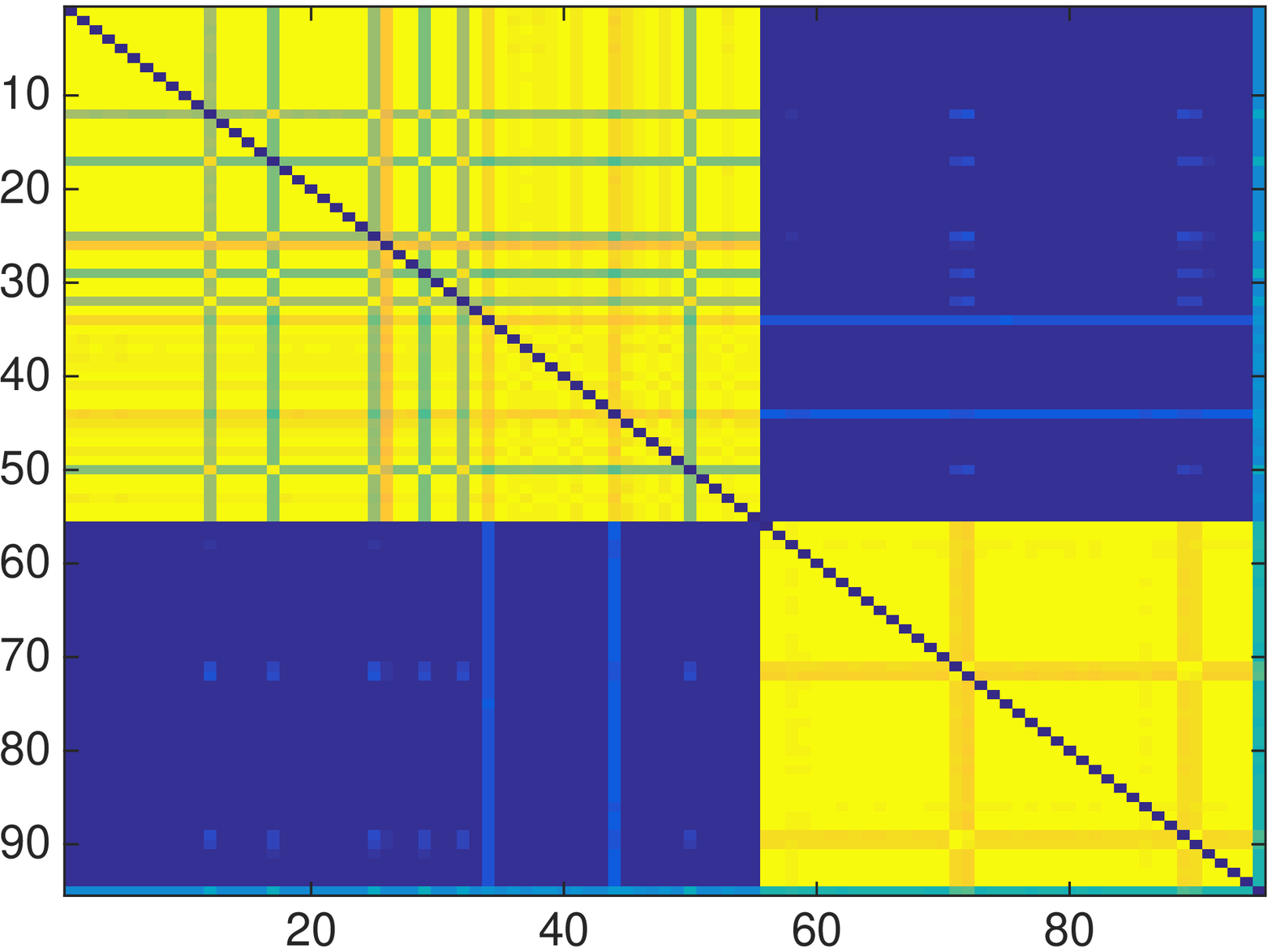} &
\includegraphics[height=1.2in]{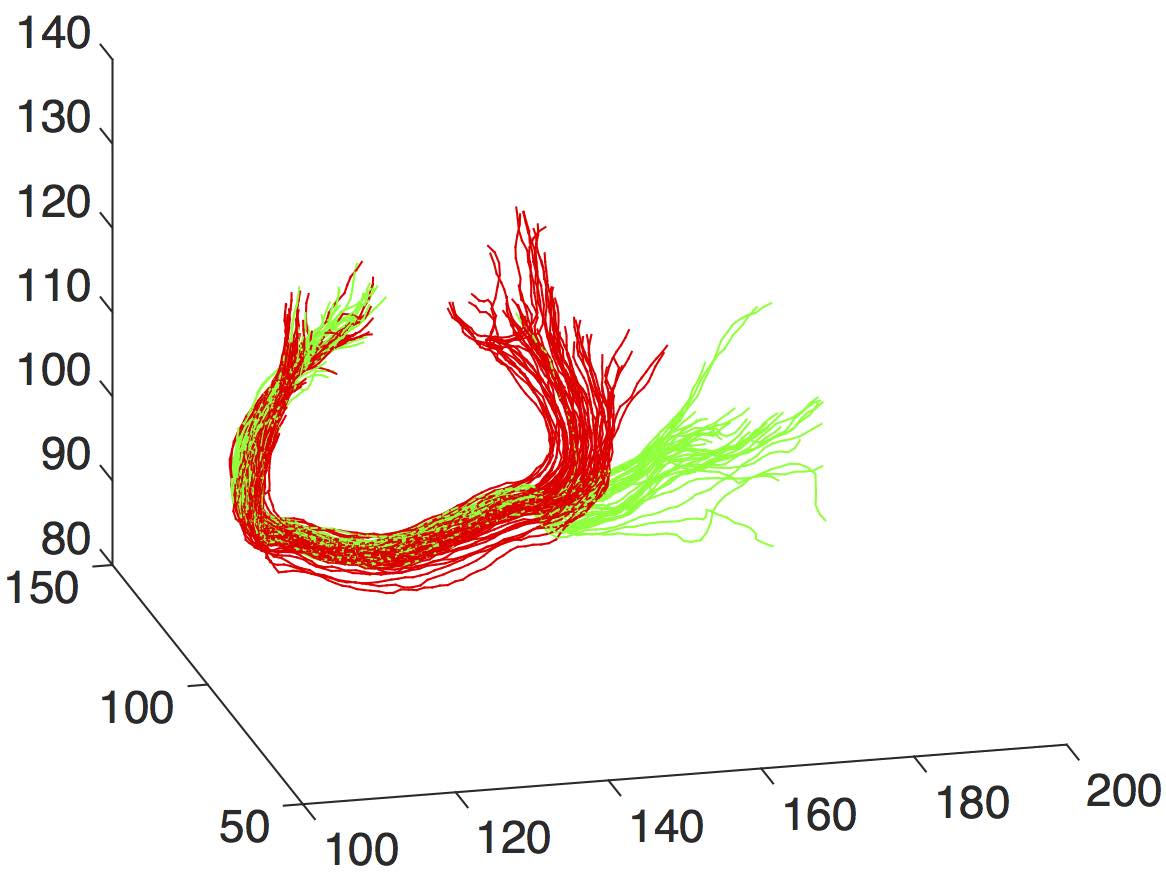} \\
(a) Posterior dist. of $k$ & (b) Adjacency matrix  & (c) Clustering on raw fibers \\
\end{tabular}
\caption{Joint model result for the connection ({\it r\_pl}, {\it l\_pg}). } \label{fig:fiber_conn1_allcomp}
\end{center}
\end{figure}

In another experiment, we applied the mixture product kernel model in (\ref{likelihood:y}) to fuse all three components together. Figure \ref{fig:fiber_conn1_allcomp} shows the results based on $30,000$ samples with a burn-in of $5,000$. From the heat map of the pairwise probability matrix, we can see that the joint mixture model prefers two clusters, and the final clustering result is similar to only using the shape component.

Similar procedures were applied to the other connection ({\it r\_pl}, {\it l\_pcc}) and the result is shown in the supplemental material. To quantitatively evaluate the modeling results, we manually clustered the fiber curves in each connection to assign ``ground truth'' labels. Fibers in ({\it r\_pl}, {\it l\_pg}) were clustered into two classes and fibers in ({\it r\_pl}, {\it l\_pcc}) were clustered into five classes (see the supplemental material for the clustering criteria and final results).  The Rand index \citep{Rand1971} and adjusted Rand index \citep{Hubert1985} are used to measure the accuracy of clustering. For any two partitions $\mathcal{C}_1$ and $\mathcal{C}_2$ of $\{1,...,n\}$, the Rand index calculates the ratio of agreement between $\mathcal{C}_1$ and $\mathcal{C}_2$ of $\{1,...,n\}$. Three quantities denoted as $a,b$ and $c$ are calculated: $a$ represents the number of pairs of objects that are placed in the same cluster in  $\mathcal{C}_1$ and the same cluster in $\mathcal{C}_2$, $b$ is the pairs that are in different clusters in both partitions, and $c$ is the total number of pairs ${n} \choose {2}$. The Rand index (RI) is $ RI = {(a+b)}/{c}$. The adjusted Rand index is corrected for chance. The Rand index can take values in $[0,1]$ and higher values indicate better agreement. The adjusted Rand index (ARI) also has a maximum value of $1$, but can yield negative values if the index is smaller than the expected index. Table \ref{tab:ri_individual} shows the quantitative result.  Again, one can confirm that the shape component contains most of the information. However, combining all components together gives us better clustering results. 

\begin{table*} \centering
\caption{Quantitative evaluation of clustering result for connections ({\it r\_pl}, {\it l\_pg} ) and ({\it r\_pl},{\it l\_pcc})}
\ra{1.0}
\begin{tabular}{@ {}rcrrrrcrrrr@{}} \toprule
& & \multicolumn{4}{c}{( {\it r\_pl}, {\it l\_pg} )} &  \phantom{abc} & \multicolumn{4}{c}{({\it r\_pl},{\it l\_pcc})} \\
\cmidrule{3-6}  \cmidrule{8-11}
  && Shape & Trans. & Rot. & All && Shape & Trans. & Rot. & All \\ \midrule
RI &&  0.8961  & 0.6090  &   0.7254  & 0.9789 &&    0.8626 &  0.8284  &  0.6767 &   0.8762 \\
ARI && 0.7923  &  0.2153  &  0.4565   & 0.9579 && 0.7088  &    0.6119 &  0.3788   &  0.7384 \\
\bottomrule
\end{tabular}
\label{tab:ri_individual}
\end{table*}

\subsection{Model for a population of individuals }
\begin{figure}
\begin{center}
\begin{tabular}{c}
\includegraphics[height=2.6in]{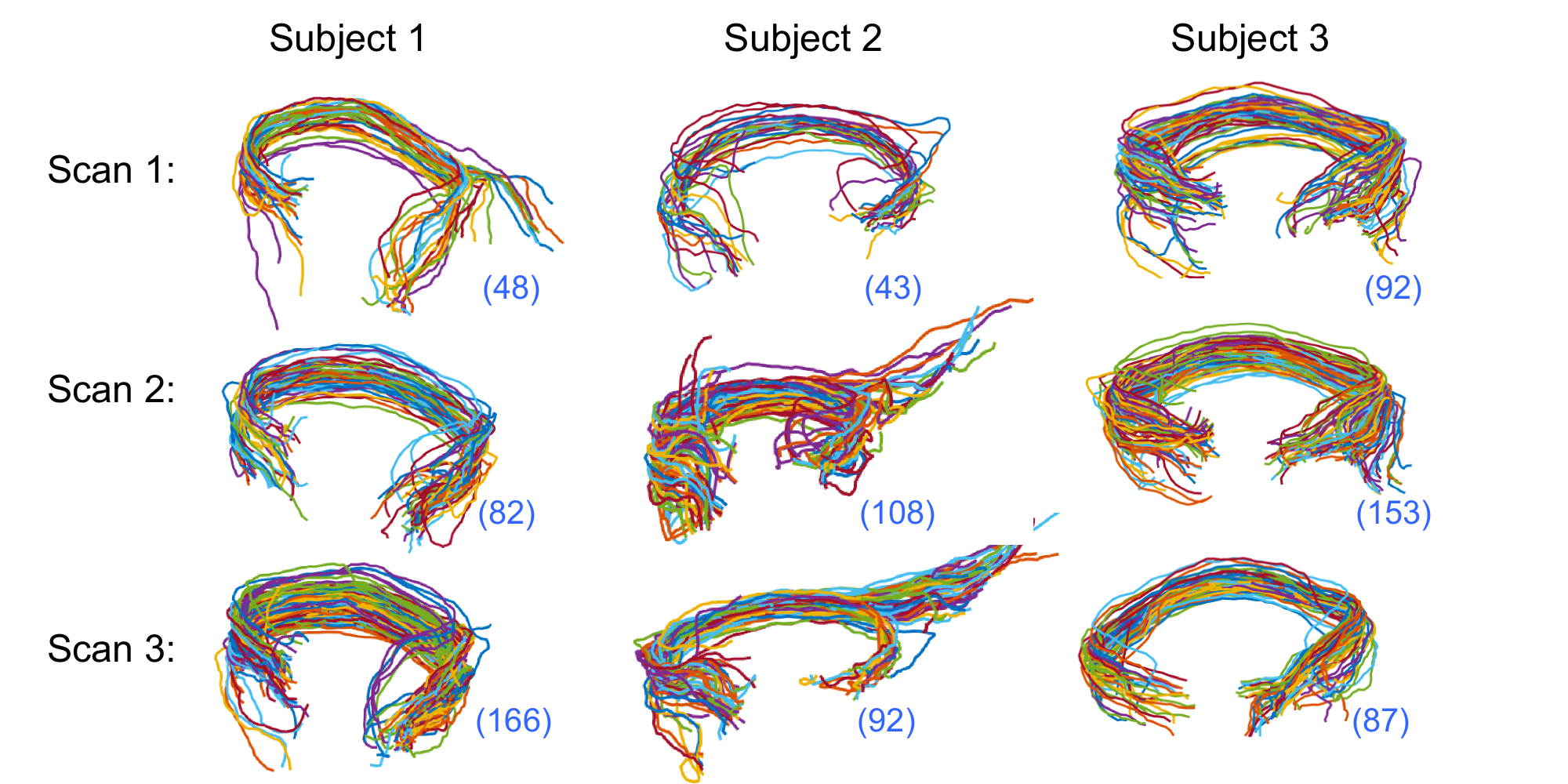}
\end{tabular}
\caption{Fiber curves connecting {\it r\_pl} and {\it l\_pg} in 9 scans of 3 subjects in the test-retest dataset.} \label{fig:roia16_roib55}
\end{center}
\end{figure}

Next, we study the connections in a set of individuals using the test-retest dataset. Figure \ref{fig:roia16_roib55} shows the fiber curves of the connection ({\it r\_pl}, {\it l\_pg}). These fibers come from three subjects in three different scans. The number in the bottom left bracket is the number of fibers in each connection.  In the routine brain network analysis literature, each connection is reduced to either a binary number ``0'' or ``1'' (to indicate whether two regions are connected) or a scalar number, e.g. the count of fibers (to indicate the strength of this connection). For  ({\it r\_pl}, {\it l\_pg}), if we reduce each connection into a binary number, there is no heterogeneity among different subjects. The rich information about the connection is totally discarded. Although one can use the count to incorporate more information about the connection, it is well known that the count of fibers can be easily contaminated  by many confounding variables in the tractography algorithm.  From Figure \ref{fig:roia16_roib55}, we can observe that for the same subject and the same connection, different scans give us different counts. The variation within subject for different scans is not smaller than variation between subjects. However, the structural connectome in healthy human brains is not expected to change rapidly across a short period. The variation of the count measure is mainly caused by the noise introduced by the tractography processing pipeline.  

An important question is whether the count can be replaced by shape information to obtain a more robust and reproducible summary of each connection; this would have significant practical ramifications in the routine analysis of brain connectome data.  By substituting in shape features, we can potentially improve the ability to 
detect differences in brain connection structure across individuals, possibly related to traits of the individual.  To assess this, we apply our NDP model to cluster individual brain connectome scans in an unsupervised manner, which does not include subject ids in the analysis.  The results in the previous section suggest that the rotation component does not contain much information, and hence we merge it into the shape part and decompose each fiber curve into two components: 
translation $c^{(1)} \in \Re^3$ and shape $c^{(2)} \in \Re^3$.  

In our first experiment, we used the fiber curves shown in Figure \ref{fig:roia16_roib55} to demonstrate our algorithm. All connections were demeaned to coarsely align them between different subjects and scans. In addition, each component was demeaned globally and rescaled to have unit variation.  These pre-processing steps simplify the prior specification. Similar to the case of modeling the fiber curves in an individual, we set $P_0 = \prod_{m=1}^M P_0^m$, where  $P_0^m \sim \text{NIW}([0,0,0]^T,1,{\bf{I}}_3,5)$, and $\alpha,\beta \sim \text{gamma}(3,3)$, a priori. The prior on $\alpha$ and $\beta$ implies that $E(\alpha) =1$ and $E(\beta)=1$, which is a common choice in the literature. The results that follow are based on $5000$ MCMC samples with a burn-in of $500$. We set $K = 9$ and $L=15$, where $K$ and $L$ are upper bounds on the number of clusters of subjects and curves within subject clusters, respectively.  

\begin{figure}
\begin{center}
\begin{tabular}{ccc}
\includegraphics[height=1.2in]{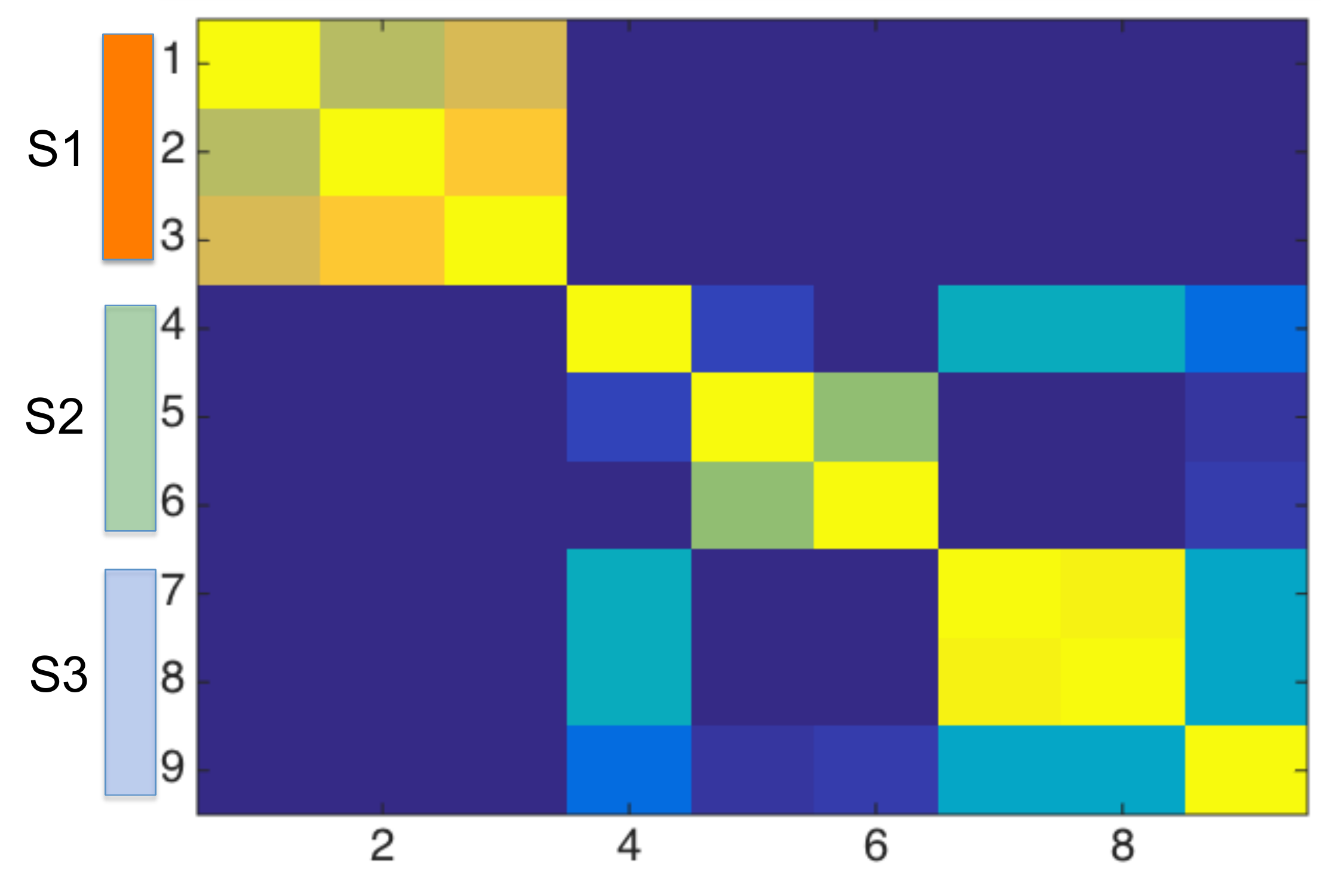}&
\includegraphics[height=1.2in]{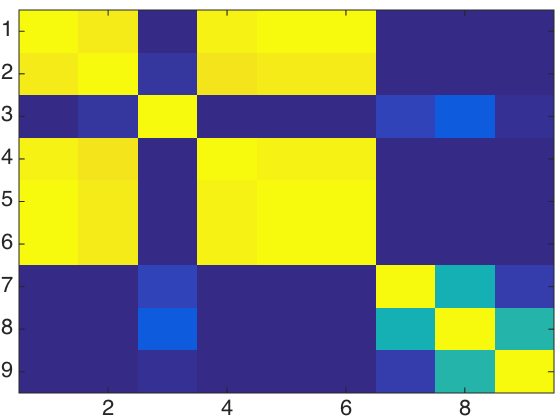} &
\includegraphics[height=1.2in]{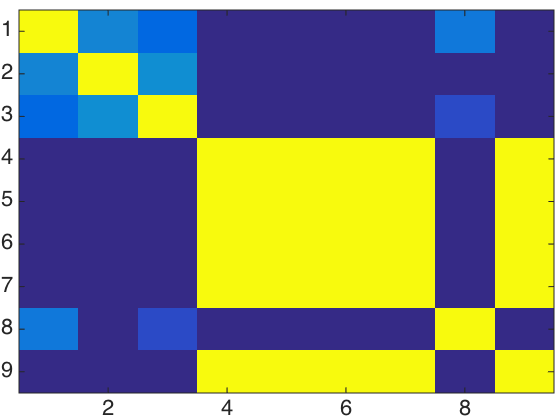} \\
(a) Shape & (b) Trans. & (c) Shape \& Trans.
\end{tabular}
\caption{Pairwise probabilities of clustering for $9$ scans of $3$ subjects in Figure \ref{fig:roia16_roib55}.} \label{fig:geo_roia16_roib55}
\end{center}
\end{figure}

Pairwise probability heat maps in different scenarios are shown in Figure \ref{fig:geo_roia16_roib55}, showing clustering results based on (a) only shape, (b) only translation, and (c) both shape and translation.  The $9$ scans were ordered by concatenating columns of Figure \ref{fig:roia16_roib55} (scans of the same subject are next to each other). From (a) we observe that, if we only use the shape part, the posterior clustering result favors five clusters: 3 scans of subject 1 are clustered together; scan 2 and scan 3 of subject 2 are clustered together and scan 1 is a separate cluster; scan 1 and scan 2 of subject 3 are clustered together and are different from scan 3. This result can be easily verified visually in Figure \ref{fig:roia16_roib55}, 3 scans of subject 1 are different from scans of subject 2 and 3 (in terms of orientation, note that these fibers are viewed from the same angle); scan 2 and 3 of subject 2 are different from scan 1; scan 1 and 2 of subject 3 are more similar comparing with scan 3. To obtain a final clustering configuration, similar to previous experiments, we  used the method in \cite{Zhang2015171} to map the pairwise probability matrix to a membership matrix.  We  compared the final NDP clustering result with the ground truth subject ids. Table \ref{tab:ri_pop} shows the Rand index and adjusted Rand index.  We can see that the shape part has the best clustering performance. Combining shape and translation does not improve clustering. 

 \begin{table*} \centering
\caption{Clustering of subjects using fiber curves connecting ({\it r\_pl}, {\it l\_pg} ).}
\begin{tabular}{@ {}r|ccccc@{}} \toprule
  && Shape & Trans. & Shape \& Trans. & Count \\
\midrule
  RI &&  0.8889  & 0.7222  &   0.7222 &0.6389\\
ARI && 0.6522  &  0.3130  &  0.3130  &-0.1818 \\
\bottomrule
\end{tabular}
\label{tab:ri_pop}
\end{table*}

\begin{figure}
\begin{center}
\begin{tabular}{cc}
\includegraphics[height=1.6in]{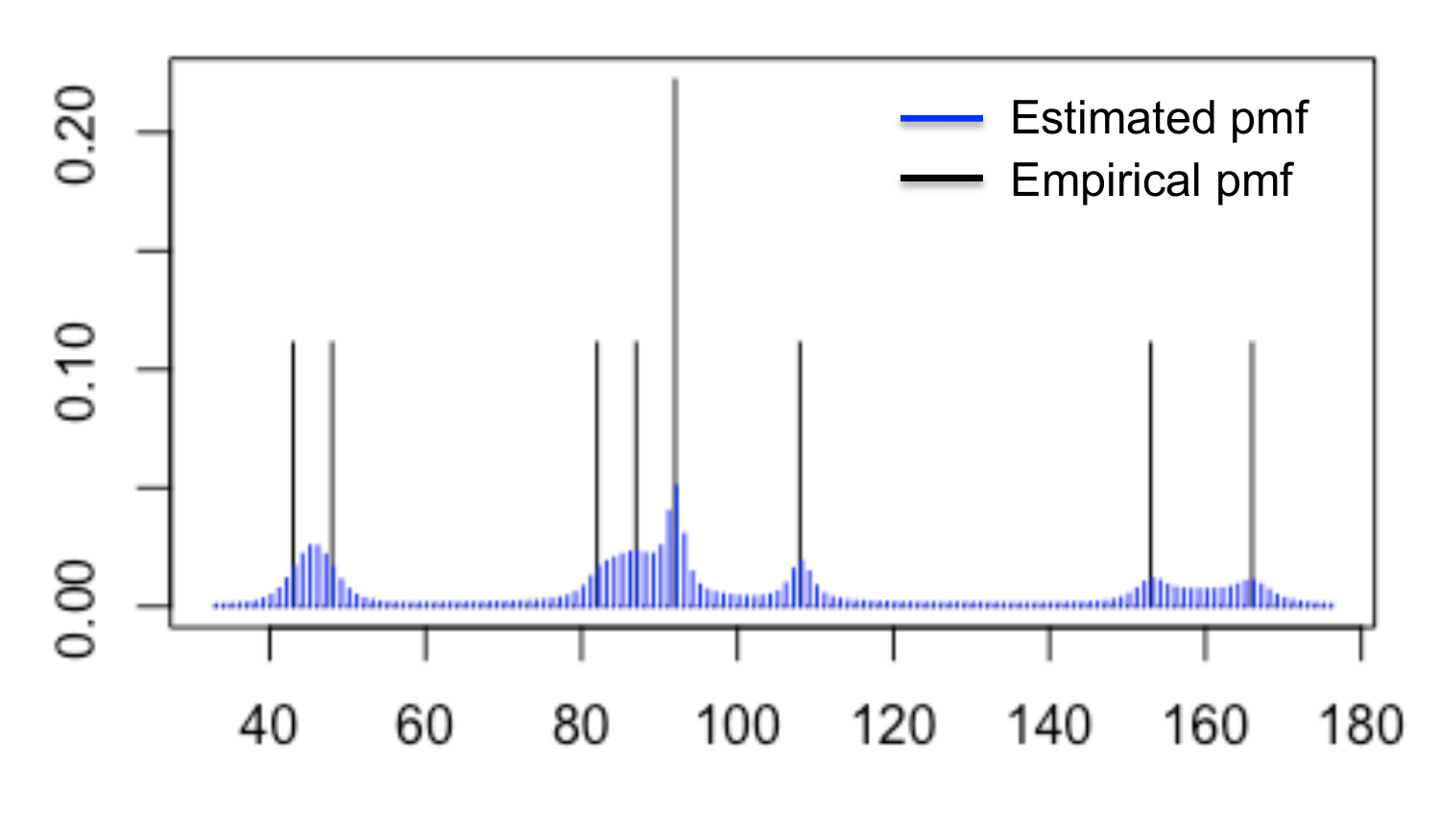}&
\includegraphics[height=1.6in]{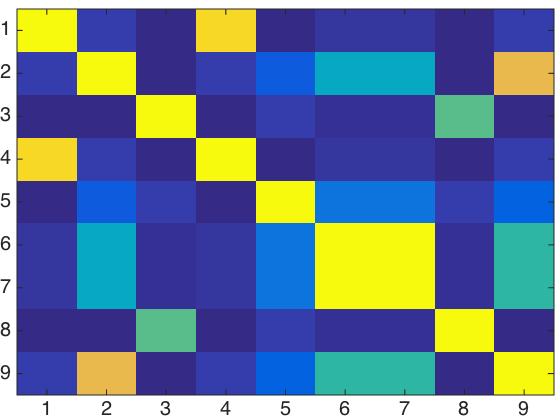} \\
(a) & (b)
\end{tabular}
\caption{(a) shows the posterior estimation of the pmf for fiber counts; (b) shows the heat map of the adjacency matrix.} \label{fig:pmf_roia16_roib55}
\end{center}
\end{figure}

As a comparison, we clustered subjects according to their fiber counts by the rounded kernel mixture model of \citet{Canale2011}, using their recommended priors, 
collecting $10,000$ posterior draws, and discarding the first $1,000$.  Figure \ref{fig:pmf_roia16_roib55} shows the result, with (a) the estimated fiber count pmf for the $9$ scans and (b) pairwise probabilities that two elements are clustered together.  The estimated pmf illustrates the enormous heterogeneity in the counts, with five peaks in the distribution.  From the pairwise probability matrix, scans for the same subject are not reliably clustered together. The Rand index and adjusted Rand index of the final clustering configuration are reported in Table \ref{tab:ri_pop}.  These results illustrate that fiber counts have very high variability and cannot reliably distinguish between subjects.

We conducted a more comprehensive analysis using all $5$ subjects and their $15$ scans from the test-retest dataset.  Connections between the left and right hemisphere having more than $30$ fiber curves were filtered out, leading to $45$ connections.  We compared clustering results using geometric information or only fiber count.  Table \ref{tab:comp_pop} shows results for $18$ connections, with the remaining results in the supplement.  The ROIs are indexed by numbers and their  names are provided in the supplement. These results provide additional evidence that shape provides the most useful summary of a connection: 
(1) shape can be reproduced robustly, (2) it is much more informative than other features (e.g., the widely used count); (3) using the whole fiber curves (shape \& translation) is not a good idea due to registration issues and the relatively limited information in the translation component.    These results suggest that future analyses of brain connectomes should ideally replace binary or count measures of connection strength with geometric features.

 \begin{table*} \centering
 \setlength{\tabcolsep}{3pt}
\caption{Comparison of clustering results of using geometric information and count.}
\begin{tabular}{r|ccccccccc} \toprule
  RI/ARI &(2,61)& (3,61) & (7,43) & (7,58) & (7,62) &(9,58) & (9,62) & (11,47)  \\
\midrule
  Shape &  0.91/0.72  & 1.0/1.0  &   0.87/0.51 &0.90/0.50& 0.90/0.47  & 0.91/0.72  &   0.91/0.72 &0.84/0.51\\
Trans. & 0.90/0.64&0.74/0.31&0.65/0.18 &0.71/0.18 &0.81/0.30&0.66/0.30 &0.60/0.16&0.52/0.15\\
  Shape \&Trans &0.70/0.3 &0.74/0.4 &0.74/0.23 &0.54/0.12&0.82/0.28&0.66/0.30&0.62/0.13&0.70/0.30\\
Count &0.64/0.23 &0.58/0.19 &0.49/0.16&0.63/0.21&0.49/0.16&0.14/0&0.45/0.01&0.58/0.19 \\
\toprule
  RI/ARI &(13,55) & (13,47) & (16,50)& (16,55)& (16,56) & (16,57) & (16,61) & (22,50)  \\
\midrule
  Shape &  0.82/0.35 & 1.0/1.0  &   0.91/0.72 &0.86/0.51& 0.83/0.53  & 0.91/0.72  &   0.90/0.64 &0.74/0.4\\
Trans. & 0.61/0.14&0.83/0.53&0.49/0.16&0.61/-0.04&0.75/0.37&0.49/0.16&0.78/0.25 & 0.58/0.19\\
  Shape \&Trans &0.70/0.21&0.74/0.40 &0.74/0.40&0.47/0.11&0.52/0.15&0.74/0.40&0.58/0.19&0.66/0.30\\
Count &0.62/0.21&0.58/0.19&0.50/0.04 &0.49/0.16 & 0.14/0&0.58/0.19&0.60/0.18&0.58/0.19\\
\bottomrule
\end{tabular}
\label{tab:comp_pop}
\end{table*}

\section{Discussion}
We have presented a novel framework to non-parametrically model the geometric information of fiber curves connecting any two brain regions.  Geometry is decomposed into three components: shape, rotation, and translation.  Our decomposition not only encourages a low dimensional representation of  the shape component but also naturally solves the misalignment issue across multiple brain scans.  Relying on a flexible hierarchical mixture model, we cluster fibers within and across individuals according to different geometric information.  These clustering results provide new insights about how to better utilize the tractography dataset for brain connectome analysis. The shape component is the most discriminative feature to distinguish different subjects and can be reliably reproduced in repeated scans. 

As a first step toward incorporating geometric information in brain structural connectome analysis, our results suggest many interesting future directions. One thread is to more intensively investigate the reproducibility of the tractography dataset from a geometric object perspective. Most previous analyses focus on analyzing arbitrarily thresholded binary networks or count weighted networks.  As we have illustrated, these features discard shape information and are highly sensitive to errors in tractography processing pipelines.  Fiber shapes appear to be significantly more robust and informative. A comprehensive study of the reproducibility of all brain connections using their geometric information can let us know which fiber bundles can be reliably reproduced.  We can assign reliability scores to every connection according to their reproducibility and give more weights to the connections with high reproducibility scores in future network analysis. This step will be fundamental in improving the reproducibility of findings in structural brain network analysis. 

Another important future direction motivated by our results is to assess the extent to which fiber tract shapes between ROIs relate to covariates and traits of the individual.  For example, neurodegenerative diseases may alter some white matter pathways and thus change the distributions of certain connections, or shapes of connections may vary systematically in relationship to cognitive abilities.  By calculating a low-dimensional set of shape features for each connection, one can obtains a ROI $\times$ ROI $\times$ feature tensor for each individual at each time point; it remains to develop appropriate statistical methods for analyzing a population distribution of such tensor-structured random variables in relation to predictors and other factors.

\section{Acknowledgments}

The authors acknowledge financial support from the Statistical and Applied Mathematical Sciences Institute and the Army Research Institute. We thank Kevin Whittingstall,  Michael Bernier, Maxime Chamberland, Gabriel Girard and Jean-Christophe Houde for acquiring the test-retest database (supported by the CHU Sherbrooke and the NeuroInformatics Research Chair jointly funded by the Medical and Science faculties).We also thank the Human Connectome Project, WU-Minn Consortium (Principal Investigators: David Van Essen and Kamil Ugurbil; 1U54MH091657) funded by the 16 NIH Institutes and Centers that support the NIH Blueprint for Neuroscience Research; and by the McDonnell Center for Systems Neuroscience at Washington University.

\appendix

\bibliography{bibfile}
\bibliographystyle{apalike}

\end{document}